\documentclass[a4paper,fleqn]{cas-dc}
\usepackage[dvipsnames]{xcolor} 
\usepackage{wrapfig,graphicx}
\usepackage{cuted}
\usepackage{epsfig,newlfont,amssymb,amsfonts,amsmath,bm,subfigure,palatino,mathtools,amsthm,braket,times,soul,enumitem,color}
\usepackage{float}
\usepackage{framed}
\usepackage[normalem]{ulem}
\newcommand{\stkout}[1]{\ifmmode\text{\sout{\ensuremath{#1}}}\else\sout{#1}\fi}
\usepackage[utf8]{inputenc}
\usepackage{array}
\usepackage{graphics}
\usepackage[
  backend=biber,
  style=numeric-comp,
  sorting=none,
  natbib=true
]{biblatex}
\DeclareFieldFormat{journalreference}{%
  \iffieldundef{url}
    {\iffieldundef{doi}
       {#1}
       {\href{https://doi.org/\thefield{doi}}{#1}}}
    {\href{\thefield{url}}{#1}}}

\DeclareBibliographyDriver{article}{%
  \usebibmacro{begentry}%
  \printnames{author}%
  \setunit{\addcomma\space}%
  \printfield[title]{title}%
  \setunit{\addcomma\space}%
  \printtext[journalreference]{%
    \printfield{journaltitle}%
    \setunit{\addspace}%
    \printfield{volume}%
    \setunit{\addspace}%
    \printtext[parens]{\printfield{year}}%
    \setunit{\addspace}%
    \iffieldundef{eid}
      {\printfield{pages}}
      {\printfield{eid}}}%
  \usebibmacro{finentry}}

\def\tsc#1{\csdef{#1}{\textsc{\lowercase{#1}}\xspace}}
\tsc{WGM}
\tsc{QE}
\tsc{EP}
\tsc{PMS}
\tsc{BEC}
\tsc{DE}

\newcommand{\ketbra}[2]{|#1\rangle \langle #2|} 
\newcommand\ddfrac[2]{\frac{\displaystyle #1}{\displaystyle #2}}

\newcommand{\ga}{\gamma_{\omega_w, w}}
\newcommand{\gb}{\gamma_{\omega_w+g, w}}
\newcommand{\gc}{\gamma_{\omega_w-g, w}}
\newcommand{\gx}{\gamma_{\omega_h, h}}
\newcommand{\gy}{\gamma_{\omega_h+g, h}}
\newcommand{\gz}{\gamma_{\omega_h-g, h}}
\newcommand{\gm}{\gamma_{\omega_c, c}}
\newcommand{\gn}{\gamma_{\omega_c+g, c}}
\newcommand{\go}{\gamma_{\omega_c-g, c}}
\newcommand{\gan}{\gamma_{-\omega_w, w}}
\newcommand{\gbn}{\gamma_{-\omega_w-g, w}}
\newcommand{\gcn}{\gamma_{-\omega_w+g, w}}
\newcommand{\gxn}{\gamma_{-\omega_h, h}}
\newcommand{\gyn}{\gamma_{-\omega_h-g, h}}
\newcommand{\gzn}{\gamma_{-\omega_h+g, h}}
\newcommand{\gmn}{\gamma_{-\omega_c, c}}
\newcommand{\gnn}{\gamma_{-\omega_c-g, c}}
\newcommand{\gon}{\gamma_{-\omega_c+g, c}}
\newcommand{\SuMo}[1]{\textcolor{black}{#1}}
\newcommand{\SuM}[1]{\textcolor{black}{#1}}



\begin{document}
\let\WriteBookmarks\relax
\def\floatpagepagefraction{1}
\def\textpagefraction{.001}
\shorttitle{Kerr-\SuMo{athermal} baths enhance cooling in quantum refrigerators}
\shortauthors{Tanaya Ray et~al.}

\title [mode = title]{Kerr-{athermal} baths enhance cooling in quantum refrigerators}                      

\author[aff1,aff4,aff2]{Tanaya Ray}
\author[aff1,aff4]{Sayan Mondal}
\author[aff1,aff4]{Aparajita Bhattacharyya}
\author[aff1,aff4,aff3]{Ahana Ghoshal}
\author[aff1,aff4]{Debraj Rakshit}
\author[aff1,aff4]{Ujjwal Sen}

\address[aff1]{Harish-Chandra Research Institute, Chhatnag Road, Jhunsi, Prayagraj  211 019, India} 
\address[aff4]{Homi Bhabha National Institute, Training School Complex, Anushakti Nagar, Mumbai 400 094, India}

\address[aff2]{School of Physical Sciences, Indian Association for the Cultivation of Science, Kolkata 700032, India}

\address[aff3]{Naturwissenschaftlich-Technische Fakult\"at, Universit\"at Siegen, Walter-Flex-Stra\ss e 3, 57068 Siegen, Germany}
\begin{abstract}
%
We study a three-qubit quantum refrigerator, with a three-body interaction enabling cooling of the target qubit, coupled to harmonic oscillator baths prepared in stationary Kerr-athermal states. The reservoir occupation probabilities are specified by a Kerr-type preparation Hamiltonian, while the bath dynamics during operation are harmonic. We show that these reservoirs lower the steady-state temperature of the cold qubit relative to the thermal-bath reference. The improvement of steady-state cooling is distinct from that of the minimum transient temperature, which is not universally enhanced. In a parameter regime where the thermal reference exhibits only transient cold-qubit cooling, the engineered Kerr-athermal baths can produce a steady-state temperature below the initial value. We also investigate the dependence of the steady-state temperature and minimum attainable temperatures on the anharmonicity in the bath state. Analysis of bath-currents  reveals lower steady-state temperatures with nearly unchanged coefficients of performance, revealing that  reservoir engineering leads to  cold-qubit cooling.
\end{abstract}





\maketitle

\section{Introduction}
\label{sec:intro}

Thermodynamics is one of the most important pillars of modern physics.
Its theoretical study has attracted attention for its fundamental importance, as also its vast applicability in everyday life.
Building thermal machines with better efficiency and tailored applicability is an ever-evolving field of scientific study and considerable progress has been made in this venture in recent years.
With the advent of quantum mechanics and the ongoing quantum technological revolution, the innovation of thermal machines suited to operate in the quantum regime has been a prime challenge for researchers.
These machines can be classified as quantum thermal machines and encompass various types, including heat engines~\cite{Scovil},  quantum refrigerators~\cite{Clivaz, Mitchison}, quantum diodes~\cite{Yuan}, quantum thermal transistors~\cite{Joulain, Zhang, Mandarino}, quantum batteries~\cite{Alicki_Fannes, Campaioli}, etc. These devices are governed by the principles of quantum thermodynamics~\cite{Kosloff2013, Uzdin2015, Goold2016, Vinjanampathy2016}. 
In the pursuit of building quantum computers suited for everyday use, we also need to provide sustainable machinery to support the operation of such highly efficient machines~\cite{Alicki_Fannes, Joulain, diode}. 
This inevitably leads to the building of thermal machines to target cooling of systems operating in the quantum mechanical scale~\cite{Linden2010, refrigerator2, Levy2012, Hewgill2020, Giazotto2006}. 

One of the most interesting among such quantum refrigerators is the self-contained ones~\cite{Skrzypczyk2011, Chen2012, Venturelli2013, Yu2014, Silva, refrigerator3, Avijit, refrigerator4, refrigerator5, refrigerator6, refrigerator7, qspeedlimit, Scarani, transient, 8fp4-15mq}, i.e.,  refrigerators that cool a quantum system without any external supply of energy. 
It has been shown that such refrigerators can be made with quantum systems as small as a qubit-qutrit {pair} or only three qubits, where one qubit is to be cooled. 
One very important aspect of such quantum machines is the use of thermal baths at different temperatures, resourcing free energy and thereby making external work redundant for quantum refrigeration.
The three-qubit quantum refrigerator, with two- or three-body interactions enabling refrigeration of the target qubit, has been studied extensively in the presence of bosonic baths composed of simple harmonic oscillators~\cite{Linden2010, Linden2010smallest, Skrzypczyk2011, Brunner2012, Levy2012, Correa2013, Correa2014,  Kosloff2014, Hewgill2020}. The paramount importance of this genre of quantum refrigerators is  
due to its faithful application in practical experimental set-ups, e.g., superconducting systems~\cite{Chen2012, Mitchison2016, Hofer2016heateng, Hofer2016, Mitchison2018}.
%
Detailed analysis involving the three-qubit quantum refrigerator have been done illustrating the impacts of various quantum resources in quantum refrigeration. Influences of quantum coherence, entanglement, and indefinite causal order have been studied in detail, and some of them have also been studied experimentally~\cite{Brunner2014, Brask2015, Mitchison2015,  Felce2020, Nie2022}. 
 These studies focus on achieving superior cooling by granting the system access to various quantum resources. Previous investigations have predominantly been done using harmonic oscillator baths, where
the thermal baths  are at suitably chosen different temperatures coupled to the { refrigerator-qubits, play a pivotal role in bringing down the temperature of the target qubit and to autonomously} 
sustain {the} operation of the machine without  aid of any external work~\cite{Linden2010}. 
It is quite evident that the nature and the state of these heat baths will play a crucial role in attaining the desired cooling capacity of such refrigerators. 

{ In the present paper, we investigate three-qubit refrigeration using harmonic oscillator reservoirs prepared in stationary Kerr-athermal states. Their occupation probabilities are defined by Gibbs weights associated with a Kerr-type preparation Hamiltonian. During the refrigerator operation, however, the bath modes evolve under harmonic Hamiltonians. The resulting states are stationary but athermal with respect to their dynamical Hamiltonians. The deformation parameter controls the departure of the occupation distribution from its harmonic thermal limit. }


{We find that the introduction of Kerr-athermal baths can lead to a considerable decrease in the steady-state temperature of the cold qubit. In Ref.~\cite{Das2019}, two parameter regimes have been identified for the refrigerator operating with thermal baths: necessarily transient-cooling regime and steady-state cooling regime. In the former, the cold-qubit temperature falls below its initial value only at transient times, while its steady-state temperature is equal to or higher than the initial value. On the other hand, in the steady-state cooling regime, the steady-state temperature is lower than the initial temperature. For the parameters studied, introducing Kerr-athermal baths lowers the steady-state temperature in both regimes. With sufficiently strong Kerr-type preparation, it falls below the initial value even in the transient-state cooling regime. The early transient minimum temperature, however, shows little or no improvement compared with the thermal-bath case. Here, a reduction of the cold-qubit temperature below its initial value is distinguished from steady-state energy extraction from the cold reservoir. The former does not by itself establish better cooling performance.}

This paper is organized as follows. In Sec.~\ref{sec:theory}, we provide the necessary theoretical framework and methodology to conduct this study. 
The concept of virtual qubit followed by the condition of refrigeration, is discussed in Sec.~\ref{subsec:virtualq}.
{In Sec.~\ref{sec:master-eq}, we introduce the stationary Kerr-athermal reservoir states and the master equation describing the refrigerator dynamics. We derive the corresponding occupation-dependent transition rates using harmonic reservoir evolution. 
In Sec.~\ref{sec:cooling}, we present the main findings of this study. We compare the transient and steady-state cooling dynamics obtained with Kerr-athermal reservoirs and conventional thermal reservoirs in Secs.~\ref{subsec:tsc} and~\ref{subsec:ssc}, respectively. In Sec.~\ref{subsec:scaling_T}, we examine how the steady-state cooling advantage depends on the Kerr deformation of the reservoir occupation distributions. Section~\ref{subsec:polar} investigates the corresponding variation in the minimum attainable temperature of the target qubit. In Sec.~\ref{subsec:cop}, we analyze the reservoir energy currents, cooling power, and coefficient of performance to distinguish improvements in the attainable temperature from changes in power and efficiency. We conclude in Sec.~\ref{conclusion}.}

\section{Autonomous quantum refrigeration in presence of {Kerr-athermal} baths}
\label{sec:theory}

\subsection{Self-contained three-qubit quantum refrigerator}
\label{sec:refg}

In this work, we consider the model of a self-contained three-qubit quantum refrigerator \cite{Linden2010} for our study. This self-contained model of quantum refrigerator allows one to cool the target qubit, called the `cold' qubit, without the necessity of any additional external work. 

The temporal behavior of the qubits is governed by the master equation describing Markovian dynamics, within the Born-Markov and secular approximations~\cite{Petruccione2002, Nielsen2011, Rivas2012}. 
The refrigerator's operation hinges on creating degeneracies within the entire three-qubit system, facilitating transitions between these degenerate states via a three-body interaction, all without any need of external energy input. { The transitions are preferentially driven by suitably chosen bath temperatures coupled to the qubits.}
The strength of this introduced interaction is very weak compared to the { relevant} energy scale of the system, so that the energy eigenstates of the composite system do not shift considerably, and hence the self-contained behaviour of such a quantum refrigerator is not compromised.

The three qubits comprising the whole system contain the `cold' qubit, i.e., the qubit to cool, and the rest of the refrigerator, plays the role of the machine. 
This machine part is made of the second and third qubits, referred to as the `hot' qubit (also called the `spiral') and the `work' qubit (labelled as the `engine'), respectively. 
The relevant parameters corresponding to the cold, hot, and work 
qubits, are labelled by the $c$, $h$, and $w$, respectively {in the rest of the article}. 
The free Hamiltonian of this three-qubit system is
\begin{equation}
    \label{eq:free-H}
    H_0 = \sum_{\alpha = c,h,w} H_\alpha = K \sum_{\alpha = c,h,w} \omega_{\alpha} \ketbra{1_{\alpha}}{1_{\alpha}},
\end{equation}
Here we denote $\ket{0_\alpha}$ and $\ket{1_\alpha}$ to be the ground and the excited states of the qubits and $K\omega_\alpha$ is the energy gap between these two energy levels, with $\alpha$ being $c$, $h$, and $w$, and the ground state energies of all the qubits have been shifted to zero. Here, $K$ is a constant having the unit of energy, and $\omega_\alpha$ are dimensionless parameters.
The excitation energies of the three qubits are chosen carefully to create the two degenerate energy states: $\ket{1_c 0_h 1_w}$ and $\ket{0_c 1_h 0_w}$, allowing the refrigerator to work in a self-contained manner. This restriction on the excitation energies leads to the following self-contained refrigeration condition: 
\begin{equation}
    \label{eq:self-contained} \omega_h = \omega_c + \omega_w .
\end{equation}
    In this self-contained (i.e., autonomous) refrigerator, the transition between the states  $\ket{1_c 0_h 1_w}$ and $\ket{0_c 1_h 0_w}$ is facilitated by the following non-local three body interaction between the subsystems
\begin{equation}
    \label{eq:int}
    H_{\text{int}} =  K g(\ket{1_c 0_h 1_w}\bra{0_c 1_h 0_w} +\ket{0_c 1_h 0_w}\bra{1_c 0_h 1_w}),
\end{equation}
where the interaction strength can be tuned by the dimensionless parameter \(g\). As mentioned before, this interaction should be weak compared to the individual energy scales of the three qubits, so {that} \(g\ll \omega_{\alpha} (\alpha = c,h,w)\). 
Hence, the total Hamiltonian of a self-contained refrigerator is given by
\begin{equation}
    \label{eq:ref} 
    H_{\text{ref}}  = H_0 + H_{\text{int}}.
\end{equation} 

In {this} self-contained refrigerator model, the three qubits are locally coupled to three thermal baths of different temperatures. 
\textcolor{black} {The state of a system, with Hamiltonian \(H_s\), at equilibrium with a thermal bath of temperature \(T\) is given by \( \rho_{\text{th}} = \exp(-H_s/k_BT)/\text{Tr}[\exp(-H_s/k_BT)]\), where $k_B$ is the Boltzmann constant. }
\textcolor{black} {Consequently,} the thermal states of the three qubits with their corresponding thermal bath's temperature ($T_\alpha$) are given by
\begin{equation}
\label{eq:thermal-dm}   
\rho_{\text{th}_\alpha} = \frac{\exp(-H_{\alpha}/k_B T_{\alpha})}{\text{Tr}[\exp(-H_{\alpha}/k_B T_{\alpha})]},
\end{equation}
with $\alpha=c,h,$ and $w$.
If we initiate the system with a three-qubit state $\rho_{\text{in}}=\rho_{\text{th}_c}\otimes \rho_{\text{th}_h} \otimes \rho_{\text{th}_w}$, where each local subsystem is diagonal in its respective energy eigenbasis (i.e., the eigenbasis of $H_\alpha$), and the refrigerator operates according to Markovian dynamics, no coherence will be generated and the local subsystems will remain diagonal in their corresponding energy eigenbasis at each successive time instant.
{\color{black} This {can be attributed to} the fact that the local Hamiltonian $H_0$ commutes with the interaction Hamiltonian, {so that} $[H_0, H_{\text{int}}]=0$, and the initial state is diagonal in the eigenbasis of $H_0$. 
{Accordingly, the time evolution of the quantum refrigerator does not couple non-degenerate energy eigenstates and thus does not generate coherence between them. 
Moreover, the Markovian evolution of each qubit precludes the generation of coherence between its corresponding local energy eigenstates across the refrigerator. 
As a result, off-diagonal elements in the global density matrix appear exclusively in the degenerate subspace, stemming from the coherence between $|010\rangle\langle 101|$ and its Hermitian conjugate.
Consequently, tracing out any two qubits leaves the reduced state of the remaining qubit diagonal in its local energy eigenbasis~\cite{Das2019}.
Since the cold qubit’s density matrix is diagonal, it can be consistently regarded as a thermal state, allowing for the introduction of a virtual temperature.
Therefore, at each instant of time, the state of any of the three subsystems can be written as}}
\begin{equation}    \rho_\alpha(t)=\rho_{\alpha}^{00}(t)\ketbra{0}{0}+\rho_{\alpha}^{11}(t)\ketbra{1}{1},
\end{equation}
where $\rho^{00}_{\alpha}(t)$ and $\rho^{11}_{\alpha}(t)$ {can be interpreted as} the occupation probabilities of the qubit in the ground state and excited state, respectively, for each $\alpha$. 
Here, $\rho^{11}_{\alpha}(t)=\exp\big(-K\omega_\alpha/k_B \tilde{\theta}_\alpha(t)\big)/\text{Tr}\big[\exp\big(-H_\alpha/k_B\tilde{\theta}_\alpha(t)\big)\big]$ and $\rho^{00}_{\alpha}(t)=\big(1-\rho^{11}_{\alpha}(t)\big)$.
\textcolor{black}{Clearly, the temperature of such a qubit can be expressed in turn, in terms of its occupation probabilities in the ground and the excited state, once it has thermalized.
Thus, the local temperatures of the qubits, defined via their occupation probabilities is }
\begin{equation}
    \label{eq:temp}
    \theta_{\alpha} = \frac{k_B \tilde{\theta}_\alpha}{K}=\frac{\omega_{\alpha}}{\ln(\rho^{00}_\alpha/\rho^{11}_\alpha)}.
\end{equation}
Here we denote $\theta_\alpha$ as the dimensionless temperatures, whereas $\tilde{\theta}_{\alpha}$ are the actual temperatures.
In order to reduce the temperature of the cold qubit, and thereby refrigerating it, one must increase its ground state population, as evident from Eq.~\eqref{eq:temp}.  
\textcolor{black}{Therefore, the refrigerator should be engineered so that the transition from the state \(\ket {1_c 0_h 1_w} \) to the state \(\ket{0_c 1_h 0_w}\) is more probable than the reverse transition, which occurs across the same energy gap. 
This is achieved by coupling the work qubit to a thermal reservoir of a much higher temperature (\(T_w\)) than that of the cold qubit, which is kept in contact with a reservoir at a lower temperature (\(T_c\)).}
The hot qubit is also maintained in contact with a thermal bath at room temperature (\(T_h\)). 
In this set-up, the initial population in the state, \(\ket {1_c 0_h 1_w} \), is much higher than the same in the state, \(\ket{0_c 1_h 0_w}\), and hence biasing the transitions within this degenerate space in the desired direction to further cool the cold qubit. 
\textcolor{black} {Under this protocol, if the subsystems begin in thermal states with appropriately chosen temperatures, the resulting heat flow leads to cooling of the cold qubit. }

\subsection{Machine virtual qubit and condition for refrigeration}
\label{subsec:virtualq}

\textcolor{black}{As outlined in the preceding section, we analyze a self-contained quantum refrigerator consisting of three qubits, each coupled to independent thermal baths at different temperatures, and subject to a tailored three-body interaction, instigating the desired cooling.
The qubit initialized at the lowest temperature, referred to as the cold qubit, is the one that is further cooled through the autonomous refrigeration process described.
However, according to the second law of thermodynamics, heat flows spontaneously from a system at a higher temperature to one at a lower temperature. To explain this seemingly counter-intuitive operation of a quantum refrigerator, Ref.~\cite{Brunner2012} introduced the concept of a `machine virtual qubit', which we briefly review in the current subsection.}

Consider the qubits $h$ (hot qubit) and $w$ (work qubit), each separately
connected to independent thermal reservoirs, and in equilibrium with their respective baths at temperatures $T_h$ and $T_w$, {to start with}. 
The two qubits interact via the three-body interaction in Eq.~(\ref{eq:int}).
\textcolor{black}{The four local energy eigenstates 
span the Hilbert space $\mathcal{H}_h \otimes \mathcal{H}_w$, in the limit of very weak interaction between the qubits.
The `machine virtual qubit'  \(v\) is defined as the subspace obtained by truncating $\mathcal{H}_h \otimes \mathcal{H}_w$  to the span of \(|0\rangle_h|1\rangle_w\) and \(|1\rangle_h|0\rangle_w\). 
Evidently, these are the two weak-coupling limit energy eigenstates that are connected by the interaction in Eq.~(\ref{eq:int}). 
Since the self-contained refrigeration condition~\eqref{eq:self-contained} dictates  \(\omega_h > \omega_w\), the ground state  is given as $|0\rangle_v := |0\rangle_h|1\rangle_w $ and the excited state  by $|1\rangle_v := |1\rangle_h|0\rangle_w $, and hence the energy gap between the two levels of the virtual qubit is $\omega_v := \omega_h - \omega_w$. Using the ratio of the populations of the two levels, the virtual temperature of this machine virtual qubit $T_v$ is embodied in the relation,
\begin{equation}
\text{exp}({-\hslash\omega_v/k_B T_v}) = \frac{p_v^1}{p_v^0} = \frac{p_h^1 p_w^0}{p_h^0 p_w^1},
\end{equation}
where $p_v^i$, $p_h^i$, and $p_w^i$ are the occupation probabilities of the $i^\text{th}$ level of the machine virtual qubit,
qubit $h$ and qubit $w$, respectively. Consequently, using the fact that the qubits are in thermal equilibrium at the temperatures of the associated heat baths, we can write,}
\begin{equation}
    \label{eq:virtual-temp}
    T_v = \dfrac{\omega_h-\omega_w}{\dfrac{\omega_h}{T_h} - \dfrac{\omega_w}{T_w}}.
\end{equation}
For specific range of values of transition energies and temperatures of the hot and work qubits, \(T_v\) can acquire values outside the range between $T_h$ and $T_w$ and can even be negative.

As discussed earlier,  the cold qubit is coupled to the hot ($h$) and the work ($w$) qubit, effectively with the machine virtual qubit, via the interaction given in~\eqref{eq:int}. The energy gap of $c$ is so chosen, such that $\omega_c = \omega_h - \omega_w$, 
which causes the states $|0\rangle_v|1\rangle_c$ and $|1\rangle_v|0\rangle_c$ to be degenerate, 
with transitions induced among the two as dictated by $H_{\text{int}}$ given in eq.~\eqref{eq:int}.
The cold qubit evolves under the influence of the local thermal bath modelling the environment at room temperature, simultaneously exchanging energy with the machine virtual qubit with virtual temperature \(T_v\).
To engineer the desired cooling of the cold qubit, the virtual temperature is made  lower than the room temperature. This modulates the evolution of the cold qubit in such a way that it approaches a steady-state temperature possibly lower than the room temperature $T_c$.  The plausible choice of
\begin{equation}
  {T_v<T_c}  
\end{equation} 
drives this evolution in the desired direction, governed by the laws of thermodynamics.
We choose the parameters for the forthcoming discussion in this work, contemplating this condition.

\subsection{Markovian master equation with \SuMo{Kerr-athermal} baths}
\label{sec:master-eq}
In this section, we discuss the details describing the system and bath working in favor of quantum refrigeration, along with the governing master equation derived to elucidate the relevant dynamics.
Let us denote the bath Hamiltonian, corresponding to the \(\alpha^{\text{th}}\) subsystem, at temperature \(T_{\alpha}\) by \(H_{B,\alpha}\) \SuMo{with 
\begin{align}
\label{eq:harmonicbath}
    H_{B,\alpha} = \sum_k \hbar \Omega_{k,\alpha} b^{\dagger}_{k,\alpha} b_{k,\alpha}.
\end{align}}
In this study, we work with \SuMo{athermal bath state where the bath is prepared in a Gibbs state at temperature $T_{\alpha}$ but with respect to the Kerr-nonlinear Hamiltonian $H_{\text{prep},\alpha}$ such that 
\begin{equation}
    \label{eq:bath-state}
    \rho_{B,\alpha} = \exp\left[-\beta_\alpha H_{\text{prep},\alpha}\right]/\text{Tr}\left(\exp\left[-\beta_\alpha H_{\text{prep},\alpha}\right]\right)
\end{equation}
where $\beta_\alpha = (k_B T_\alpha)^{-1}$ is the inverse temperature. The Hamiltonian $H_{\text{prep},\alpha}$ is given by
 \begin{equation}
  \label{eq:bath}    
    H_{\text{prep},\alpha} = \sum_k \hbar \left(\Omega_{k,\alpha} b^{\dagger}_{k,\alpha}b_{k,\alpha} + \frac{\Omega_{k,\alpha}}{\zeta}  (b_{k,\alpha}^{\dagger}b_{k,\alpha})^2 \right),
\end{equation}
where \(k\) runs over the infinite number of bath modes and the dimensionless parameter \(\zeta\) modulates the relative energy-shift of a bath oscillator due to its anharmonicity, compared to the harmonic case.
$\Omega_k^{\alpha}$ represents the frequency of the $k^{\text{th}}$ mode of the $\alpha^{\text{th}}$ bath, and $b_{k,\alpha}^\dagger(b_{k,\alpha})$ is the bosonic creation(annihilation) operator obeying the commutation relation $[b_{k,\alpha},b^\dagger_{l,\alpha}]=\delta_{kl}$. }
\textcolor{black}{Such anharmonicity potentially models effective repulsive interaction of bosons trying to occupy a single bath mode. 
Influenced by such interaction, the spacing between two consecutive number states increases with increasing number. }
The described shifts in energies can also be explained by Kerr-type nonlinearity and hence the oscillators utilized can be called Kerr-type oscillators. Evidently, the \(\zeta \rightarrow \infty\) limit in Eq.~\eqref{eq:bath} gives back the usual harmonic oscillator bath in Eq.~\eqref{eq:harmonicbath}.
{It is also worth noting in this context that anharmonicity is an indispensable feature of quantum systems, as it enables precise control over their operation and excitations. This is what makes transmons and systems with Kerr-type anharmonicity the cornerstone elements of modern quantum devices \cite{Gu2017}.}

\SuM{The free bath evolution used to evaluate the reservoir correlation functions is generated by \(H_{B,\alpha}\). Since \(H_{\mathrm{prep},\alpha}\) and \(H_{B,\alpha}\) are functions of the same number operators, \([\rho_{B,\alpha},H_{B,\alpha}]=0\), and the prepared state is stationary under this evolution. For finite positive \(\zeta\), however, it is not a Gibbs state of the operating Hamiltonian \(H_{B,\alpha}\). The parameter \(T_\alpha\) specifies the preparation distribution and is chosen to coincide with the initial temperature of qubit \(\alpha\); it is not an equilibrium temperature of the operating reservoir. The harmonic Gibbs state at \(T_\alpha\) is recovered as \(\zeta\to\infty\).}

\textcolor{black} {We consider the system–bath coupling to be weak enough to validate the Born–Markov and secular approximations, in support of using the relevant master equations}~\cite{Petruccione2002, Nielsen2011, Rivas2012}. 
The interaction between qubit \(\alpha\) and its \SuMo{Kerr-athermal} bath \textcolor{black}{in the considered quantum refrigerator model} is governed by the Hamiltonian 
\begin{align}
    \label{eq:sys-bath}
    H_{SB,\alpha} 
    =\sum_k \hbar \lambda_{k,\alpha} (\sigma_{\alpha}^+ b_{k,\alpha} + \sigma_{\alpha}^- b_{k,\alpha}^{\dagger} ), 
\end{align}
where \(\lambda_{k,\alpha}\), having the unit of frequency,  stands for the control parameter to determine the strength of interaction between a qubit and the $k^{\text{th}}$ mode of the corresponding thermal bath. 
Also, we assume $\lambda_{k,\alpha}^2=\frac{K}{\hbar}
J_\alpha(\Omega_k)$, where $J_\alpha(\Omega_k)$ is the spectral density function of the $\alpha^{\text{th}}$ bosonic bath. 
\(\sigma_{\alpha}^+ (\sigma_{\alpha}^-)\) are the raising(lowering) operators of the qubit \(\alpha\).
So, the total Hamiltonian, including the system, all the baths locally connected to the three qubits, and all kinds of interactions relevant to our purpose, is expressed as 
\begin{equation}
    \label{eq:H_total}
    H_T = H_{\text{ref}} + \sum_{\alpha = c,h,w} H_{B,\alpha} + \sum_{\alpha = c,h,w} H_{SB,\alpha}.
\end{equation}
Under the Born-Markov and secular approximations, the density matrix of the three-qubit system follows 
the Gorini-Kossakowski-Sudarshan-Lindblad (GKSL) master equation \cite{Petruccione2002, Nielsen2011, Rivas2012}, 
and can be expressed as 
\begin{equation}
    \label{eq:QME}
    \frac{d\rho_s(t)}{dt} = - \frac{i}{K} [H_{\text{ref}},\rho_s(t)] +\frac{\hbar}{K} \sum_{\alpha = c,h,w} \mathcal{D_{\alpha}}(\rho_s(t)),
\end{equation}
where the dissipator \(\mathcal{D_{\alpha}}(\rho_s(t))\) takes the form
\begin{equation}
\label{eq:D}
\begin{split}
    & \mathcal{D_{\alpha}}(\rho_s(t))  \\ 
    & = \sum_{\mathcal{E}} \gamma_{\mathcal{E},\alpha} ( A_{\mathcal{E},\alpha} \rho_s(t) A_{\mathcal{E},\alpha}^{\dagger}  - \dfrac{1}{2}\{ A_{\mathcal{E},\alpha}^{\dagger}A_{\mathcal{E},\alpha}, \rho_s(t) \} ). 
\end{split}
\end{equation}
Here, $t$ denotes the dimensionless time parameter. 
The Lindblad operator \(A_{\mathcal{E},\alpha}\) is responsible for transitions 
between the eigenstates of the Hamiltonian \(H_{\text{ref}}\), separated by the transition energy \(\hbar\mathcal{E}\). The summation in Eq.~\eqref{eq:D} is over all such distinct transition energies obtained from the set of eigenstates of the Hamiltonian \(H_{\text{ref}}\). The Lindblad operators are basically the system operators, 
expressed in the eigenbasis of the total system Hamiltonian \(H_{\text{ref}}\). For the system-bath interaction as in Eq.~\eqref{eq:sys-bath}, the Lindblad operators take the form \cite{TRay2023}
\begin{equation}
    \label{eq:A-op}
    A_{\mathcal{E},\alpha} = \sum_{e_k-e_l=\hbar\mathcal{E}} \ketbra{e_l}{e_l} \sigma^x_{\alpha} \ketbra{e_k}{e_k},
\end{equation}
\(\ket{e_i}\) being the energy eigenstate of \(H_{\text{ref}}\) with energy \(e_i\). 
\textcolor{black}{The 
decay rates,
\(\gamma_{\mathcal{E},\alpha}\), in Eq.~\eqref{eq:D}, are proportional to the real part of the power spectrum obtained via one-sided Fourier transform of the bath correlation function.}
Also, the  local heat baths are assumed to act independently, 
so that we can write the total Lindbladian as the sum of the dissipators arising from each bath as expressed in Eq.~\eqref{eq:QME}. 
\textcolor{black}{It is important to note that in the original work of Linden et al.~\cite{Linden2010}, the inter-qubit interaction $g$ was considered to be negligible, and the cooling mechanism was explained using population biases introduced due to the interaction. In subsequent works~\cite{Correa2013,Correa2014, Das2019, Brask2015, Levy2012}, the global master-equation approach was adopted where the dissipative dynamics were derived in the eigenbasis of the full interacting system Hamiltonian. In this work we do the same, as it is more accurate and it provides a more consistent microscopic description when the interaction between subsystems is not negligible. Moreover, it allows us to
capture cooling features that may be neglected within a purely local description. Hence, this type of
master equation is like a dressed master equation.} Under the Born-Markov approximation, such decay rates can be 
simplified to the expression
\begin{equation}
    \label{eq:gamma-1} 
      \gamma_{\mathcal{E},\alpha}=\left\{
    \begin{array}{ll}
      \frac{\pi}{2} J_{\alpha} (\mathcal{E})[\langle n_{\beta}(\mathcal{E}) \rangle +1], & \mbox{if $\mathcal{E}>0$}.\\
      \frac{\pi}{2} J_{\alpha} (|\mathcal{E}|) \langle n_{\beta}(|\mathcal{E}|) \rangle , & \mbox{if $\mathcal{E}<0$}.
    \end{array}
  \right.
\end{equation}
Here, \(\langle n_{\beta}(\mathcal{E}) \rangle \) is the 
average particle number at energy $\hbar \mathcal{E}$, in the thermal state of the corresponding bath at inverse temperature \(\beta = 1/k_BT_{\alpha}\).
For our purpose, we assume the bath spectral density to be faithfully captured by that of an Ohmic bath, given by
\begin{equation}
    \label{eq:j-w}
    J_{\alpha} (\mathcal{E}) = \frac{2}{\pi} \delta_{\alpha} \frac{|\mathcal{E}|}{\omega_{\alpha}} \exp(-|\mathcal{E}|/\Omega_c),  
\end{equation}
where 
\(\Omega_c\) is the cut-off frequency to avert the divergence of such Ohmic bath spectral density at frequencies very large compared to the energy scale of the system and $\delta_\alpha$ is a constant having the unit of frequency. In the entire discussion of this paper, we fix
\(\Omega_c=5000 \frac{K}{\hbar}\), which provides us a sensible choice to taper off the \(J_{\alpha} (\mathcal{E})\) at very high \(\mathcal{E}\) values. Moreover, for the validation of Born-Markov approximations, the coupling between the qubits and the thermal baths should be very weak 
and hence we limit the parameter space such that \(\gamma_{\mathcal{E},\alpha} \ll |\mathcal{E}|\) for all relevant transition energies \(\hbar \mathcal{E}\). \textcolor{black}{The explicit forms of the Lindblad operators for this refrigerator model and the steps for obtaining the steady-state are given in Appendix~\ref{appen:B}.}

In the following derivation in this subsection, we have dropped the implied \(\alpha\) index associated with various parameters in spirit of making the expressions look cleaner.
Here we compute the \SuMo{mean occupation number in different bath modes contributing to the decay rates as expressed in Eq.~\eqref{eq:gamma-1}, for Kerr-athermal bath states} described in Eq.~\eqref{eq:bath}. This average particle number comes out as

\begin{align}
    \label{eq:n_av}
     \braket{n_{\beta}(\mathcal{E})}
    = \ddfrac{\sum\limits_{n_{\mathcal{E}}=0}^{\infty}n_{\mathcal{E}}\exp\Big[-\hbar\beta\left(\mathcal{E} n_{\mathcal{E}}  + \frac{\mathcal{E}}{\zeta}  n_{\mathcal{E}}^2 \right)\Big]}{\sum\limits_{n_{\mathcal{E}}=0}^{\infty}\exp\Big[-\hbar\beta\left(\mathcal{E} n_{\mathcal{E}}  + \frac{\mathcal{E}}{\zeta}  n_{\mathcal{E}}^2 \right)\Big]}.
\end{align}
See Appendix~\ref{appen:A} for the detail calculation of $\braket{n_{\beta}(\mathcal{E})}$. The series in the above expression is not analytically computable, 
though the integral test ensures their convergence. So,
to compute the 
decay rates in the master equation (Eq.~\eqref{eq:QME}), we use Eq.~\eqref{eq:n_av} and numerically calculate the average number of particles in the corresponding bath temperature and transition frequencies.

As discussed earlier, in the regime, $\zeta \rightarrow \infty$, we obtain the usual thermal bath. Consequently, in this limit, Eq.~\eqref{eq:n_av} reduces to,
\begin{equation}
    \begin{split}
        \langle n_\beta(\mathcal{E})\rangle & = 
        \ddfrac{\sum\limits_{n_{\mathcal{E}}=0}^{\infty}n_{\mathcal{E}}\exp(-\hbar\beta\mathcal{E} n_{\mathcal{E}})}{\sum\limits_{n_{\mathcal{E}}=0}^{\infty}\exp(-\hbar\beta\mathcal{E} n_{\mathcal{E}})}= \frac{1}{e^{\hbar\beta \mathcal{E}} - 1},
    \end{split}
\end{equation}
which is just
the well-known Bose-Einstein distribution function, as expected~\cite{Rivas2012}.

\begin{figure*}[!!t]
\includegraphics[width=\textwidth]{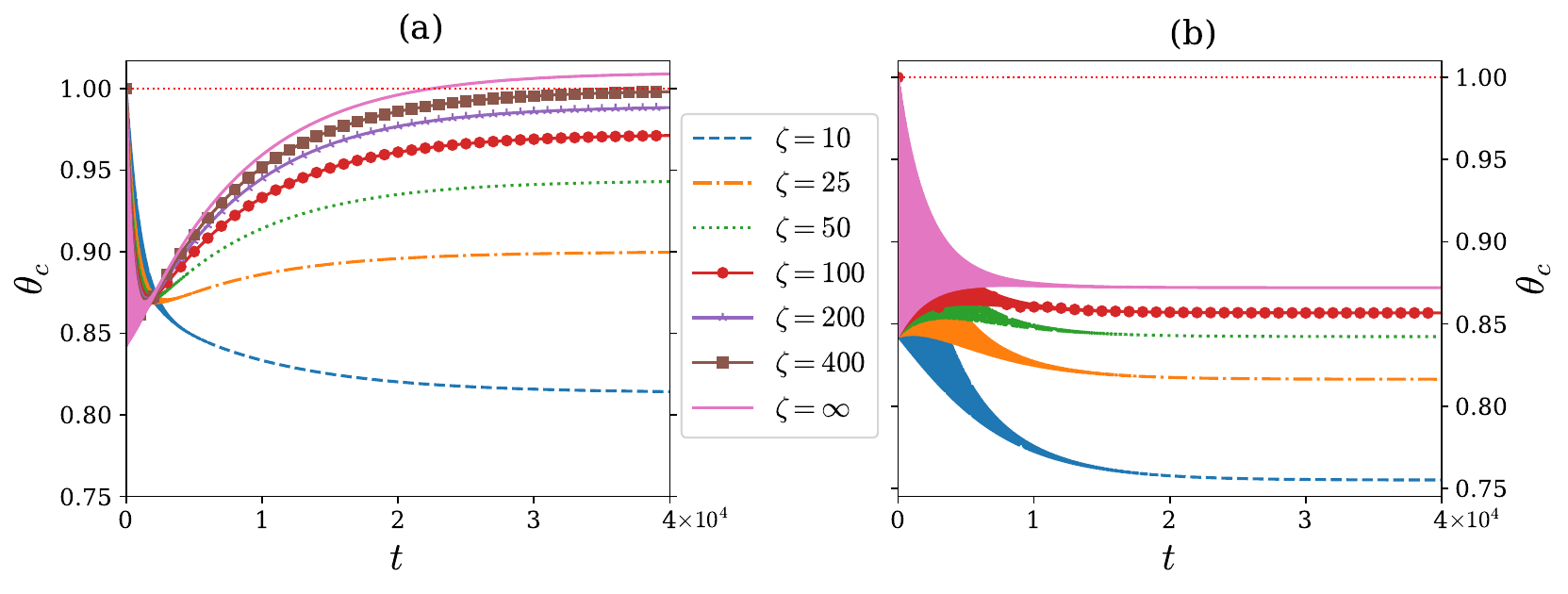}
\caption{Time evolution of the cold qubit temperature (\(\theta_c\)) by varying the \SuMo{anharmonicity of the preparation Hamiltonian $H_{\text{prep},\alpha}$ of the Kerr-athermal baths.} 
Here we plot the 
time dynamics of \(\theta_c\) with varying level of anharmonicities $(\zeta)$ in \SuMo{$H_{\text{prep},\alpha}$}. Increasing the value of \(\zeta\) signifies decreasing anharmonicity, reaching the \SuMo{thermal} case in the limit \(\zeta\rightarrow\infty\). The initial temperature of the cold qubit is marked by the red dotted lines. 
Panel (a) describes the variation of \(\theta_c\) when the system parameters are chosen to obtain transient cooling in the presence of \SuMo{thermal} baths, as specified in Sec.~\ref{subsec:tsc}. 
Panel (b) shows the time dynamics of \(\theta_c\) in the case where steady-state cooling is achieved 
{\color{black}for the parameters which provide steady-state cooling in the \SuMo{thermal} bath case.} 
The relevant parameters are mentioned in Sec.~\ref{subsec:ssc}. All the quantities plotted here are dimensionless.
}
\label{fig:tc-ssc}
\end{figure*}

\section{Enhanced cooling with \SuMo{Kerr-athermal} baths}
\label{sec:cooling}

\textcolor{black}{We now present the autonomous refrigeration of a three-qubit refrigerator in the presence of \SuMo{Kerr-athermal} baths and describe the advantage of using such baths to cool a quantum system compared to the scenario when \SuMo{thermal baths}  are used.
 In previous works~\cite{Linden2010, Levy2012, Correa2014, Brask2015, Mitchison2015, Das2019}, it was shown that we can choose the parameters \(\{\delta_{\alpha} | \alpha \in\{\textit{c}, \textit{h}, \textit{w}\}\}\) (see Eq.~\eqref{eq:j-w}) from different parameter sub-spaces, depending on the choice of which, one can fine-tune the dynamics of the temperature attained by the cold qubit in the three-qubit quantum refrigerator.}
 Throughout the following discussion, we distinguish two such distinct parameter-spaces as $R_T$ and $R_S$.
 With $R_T$ we denote the regime where the chosen parameters yield us with only transient cooling of the cold qubit, followed by its heating in steady state, when the employed baths are thermal. 
 \textcolor{black}{In this regime, the anharmonicity in the preparation Hamiltonian $H_{\text{prep},\alpha}$ of the bath modes leads to better steady-state cooling than in the \SuMo{thermal} baths case. Better transient-regime temperatures are also attained, even if the minimal temperature reached is often in the steady state.}
 $R_S$ specifies the parameter regime where both transient and steady-state cooling is achieved using thermal baths. \SuMo{Kerr-athermal} baths come out to be favorable in this case, whether one targets to achieve better steady-state cooling or the lowest transient temperature. We subsequently describe the time dynamics of the temperature of the cold qubit in both of these cases and demonstrate the usefulness of using \SuMo{Kerr-athermal} baths instead of \SuMo{thermal} ones. 

For the subsequent investigations, we fix 
$\omega_c=\omega_w=K/\hbar$ and 
\(\omega_h=2K/\hbar\). The parameters $T_\alpha$ in the $H_{\text{prep},\alpha}$ of both the heat baths thermalizing the cold and hot qubit are taken to be the same: \(T_c=T_h = K/k_B \). The work qubit remains in contact with a bath of \(T_w=2 K/k_B\) throughout. 
We initialize the three qubits in the state $\rho_\text{in} = \rho_{\text{th}_c}\otimes\rho_{\text{th}_h}\otimes\rho_{\text{th}_w}$, where $\rho_{\text{th}_\alpha}$ is given in Eq.~\eqref{eq:thermal-dm} with the respective $T_{\alpha}$. \SuMo{The choice of this initial state is motivated by the fact that we are comparing the Kerr-athermal baths with the thermal baths and their ability to cool the cold qubit. Hence, in both the cases, we start with identical initial states of the system.}
Following the discussion in Sec.~\ref{subsec:virtualq}, we can see that these choices of parameters facilitate the desired refrigeration.

\subsection{Regime $ R_T $:  anharmonicity improves steady-state cooling but not  minimal transient temperature} 
\label{subsec:tsc}

\textcolor{black}{To scrutinize the three-qubit quantum refrigerator in the parameter space where we get transient cooling and steady-state heating using \SuMo{thermal} baths, we choose the qubit-bath coupling parameters as \( \delta_c=10^{-4}\omega_{c}\), \(\delta_h=10^{-5}\omega_{h}\), \(\delta_w=10^{-3}\omega_{w} \). 
The interaction strength facilitating the autonomous cooling is fixed by the parameter \(g\) in Eq.~\eqref{eq:int}, which is chosen to be \(0.8\) in this case. As can be seen from the \(\zeta \to \infty\) (thermal limit of the baths) plot in Fig.~\ref{fig:tc-ssc}(a), the chosen set of 
parameters indeed facilitates transient cooling of the cold qubit at small times, although there is heating of the same qubit in the steady state.
From the same figure, we can conclude that as we introduce anharmonicity in the \SuMo{preparation Hamiltonians $H_{\text{prep},\alpha}$}, not only does the steady-state temperature of the target qubit decrease but also it opens up the scope of steady-state cooling of the cold qubit in the same parameter regime where the presence of \SuMo{thermal} baths failed to cool it in the steady state. 
Smaller values of \(\zeta\) contribute to larger anharmonicity in the bath modes of \SuMo{$H_{\text{prep},\alpha}$}, and larger values of \(\zeta\) contribute to smaller anharmonicity, leading to the \SuMo{thermal case} in the limit \(\zeta \rightarrow\infty\). As expected, and can be verified from the Fig.~\ref{fig:tc-ssc}(a), larger anharmonicity facilitates better steady-state cooling of the cold qubit, though it takes a very small amount of anharmonicity (\(\zeta\lesssim 400\)) to achieve steady-state cooling in the chosen parameter space.
As explored in some previous works, the cold qubit attains a much smaller temperature in the transient regime before it gets heated (with respect to the transient state temperatures) and reaches the steady state. 
From Fig.~\ref{fig:tc-ssc}(a), we can see that though introducing anharmonicity in the \SuMo{$H_{\text{prep},\alpha}$} helps us to cool the cold qubit more efficiently in the steady state,  the relative advantage of attaining the minimal transient temperature in a very short time diminishes very slowly as the bath move further from the \SuMo{thermal} regime. From Fig.~\ref{fig:tc-ssc}(a), it is also clear that the feature of having a lower transient temperature than the steady-state one happens for \(\zeta \gtrsim 25\).} 
It may be noted that for short times, no amount of anharmonicity can triumph over the \SuMo{thermal} bath case, in terms of the lowest temperature achieved. As can be seen below in the next subsection, this is no longer true for the Regime \(R_s\). 

\subsection{Regime $R_S$: anharmonicity improves cooling 
at all times}
\label{subsec:ssc}

 \textcolor{black}{In this parameter regime, the three-qubit quantum refrigerator can incorporate steady-state refrigeration, as well as lower transient temperatures of the cold qubit in the presence of local thermal heat baths. 
 We study the same refrigerator, but in the presence of \SuMo{Kerr-athermal baths}, in this regime with a fixed set of parameters: interaction strength between the subsystems \(g=0.1\), and uniform qubit-bath interactions \( \delta_c=10^{-4}\omega_{c}\), \(\delta_h=10^{-4}\omega_{h}\), \(\delta_w=10^{-4}\omega_{w} \). 
The \(\zeta \to \infty\) case (i.e. for thermal baths) in Fig.~\ref{fig:tc-ssc}(b) clearly shows that the chosen set of parameters leads to a steady-state cooling of the cold qubit, and  even lower transient temperatures. 
%
With incorporation of anharmonicity in \SuMo{$H_{\text{prep},\alpha}$}, the steady-state temperature goes further down.
 The cold qubit attains a lower temperature in the transient regime  than that in the steady state, even 
 for small anharmonicities. 
 However, with stronger anharmonicity in the bath athermal state, viz. \(\zeta \lesssim 50\),
  the effect 
 vanishes.
%
This aspect is discussed in further detail in Sec.~\ref{subsec:polar} of this paper. }

\begin{figure}
 \includegraphics[width=0.5\textwidth]{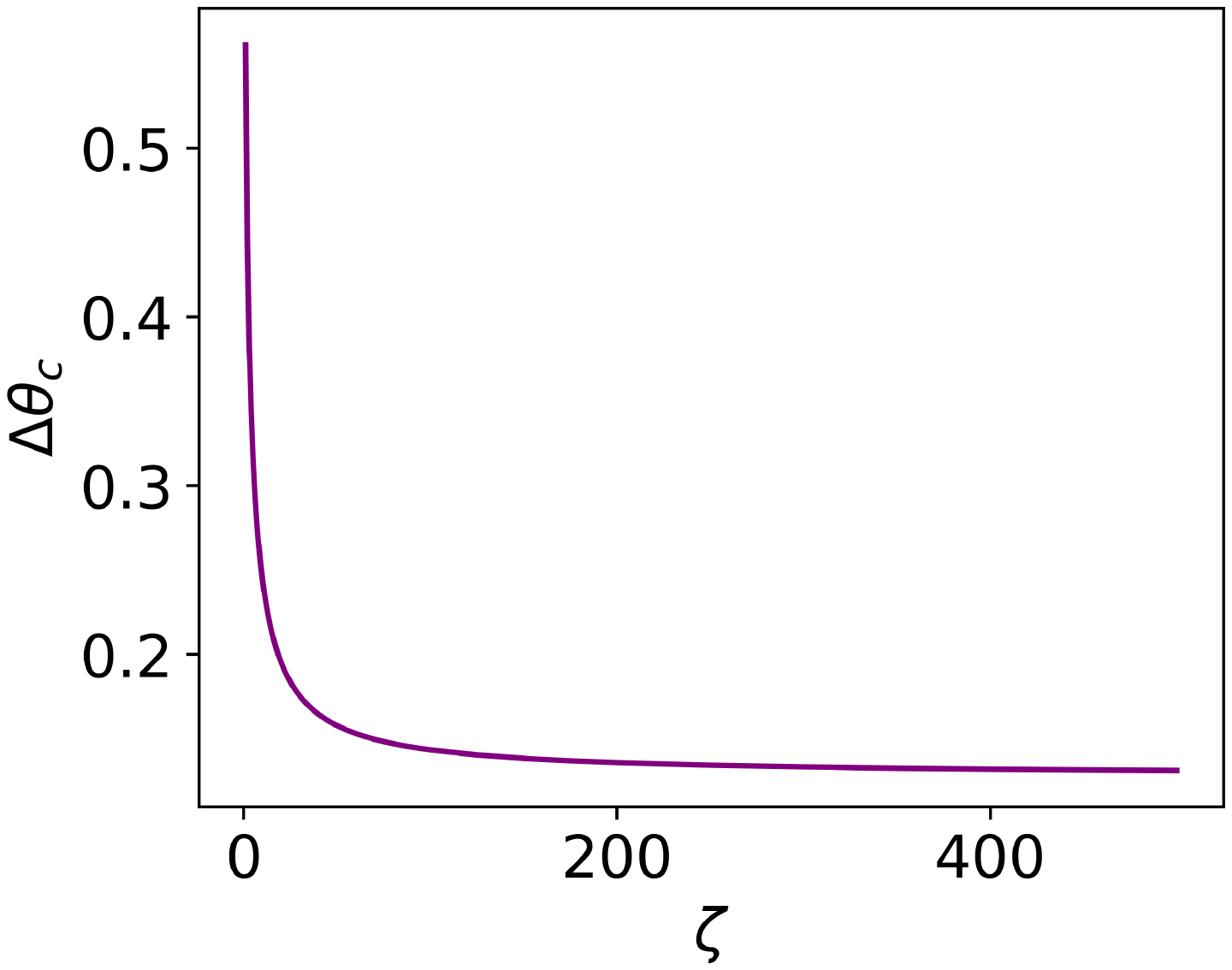}
\caption{Variation of the steady-state cooling advantage with varying anharmonicity in the \SuMo{preparation Hamiltonian $H_{\text{prep},\alpha}$ of the Kerr-athermal baths}. \(\zeta\) captures the mentioned anharmonicity as described by Eq.~\eqref{eq:bath} and is plotted along the horizontal axis. Plotted along the vertical axis is \(\Delta\theta_c\), the corresponding steady-state temperature of the cold qubit subtracted from its initial value. All the relevant parameters are taken as described in Sec.~\ref{subsec:ssc}. All the quantities plotted here are dimensionless.
}
\label{fig:scaling_dT}
\end{figure}
\begin{figure}
 \includegraphics[width=0.5\textwidth]{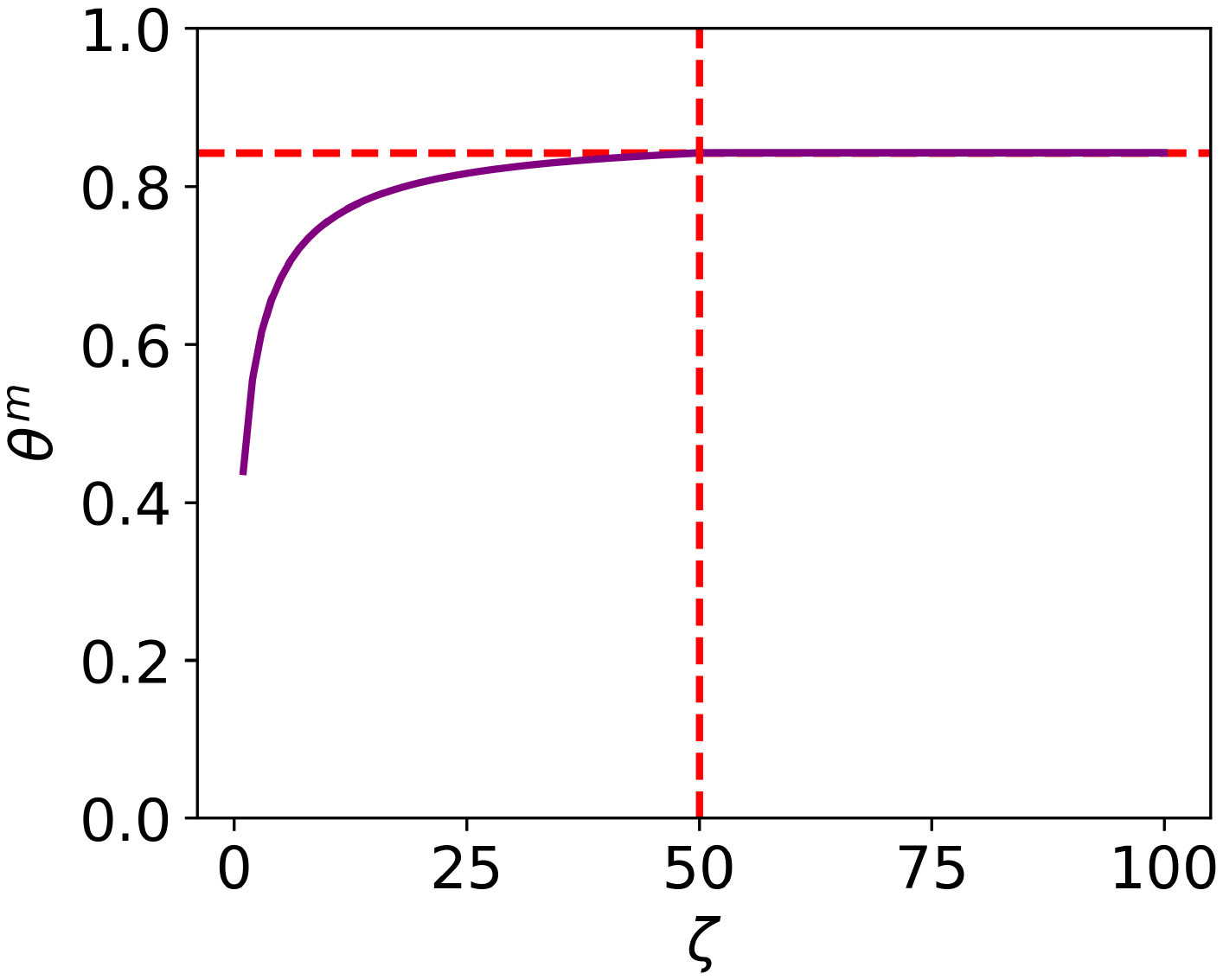}
\caption{Variation of the minimum attainable temperature of the cold qubit with changing anharmonicity of the \SuMo{preparation Hamiltonian $H_{\text{prep},\alpha}$ of the Kerr-athermal baths}. \(\zeta\) depicts the anharmonicity according to Eq.~\eqref{eq:bath} and is plotted along the horizontal axis. \(\theta^m\) is the corresponding global minimum of temperatures the cold qubit evolves through, plotted along the vertical axis. All the relevant parameters are mentioned in Sec.~\ref{subsec:ssc}. All quantities plotted here are dimensionless.
}
\label{fig:polar-temp}
\end{figure}

\SuM{
\subsection{Physical intuition behind the decrease in steady-state temperature}
The physical mechanism behind the decrease in steady-state temperature can be understood by examining how the Kerr-type preparation modifies the bath occupation statistics and the associated transition rates.
In the refrigerator based on the Kerr-athermal baths, for a mode of frequency $\nu$ in bath $\alpha$, the ratio of successive occupation probabilities is
\begin{align*}
    \frac{p_{n+1}}{p_n} = \exp\left[-\beta_\alpha \hbar \nu \left(1 + \frac{2n+1}{\zeta}  \right)\right].
\end{align*}
This ratio is smaller for finite $\zeta$ values, when compared to the thermal baths with $\zeta\to \infty$ at same value of $T_\alpha$. 
Thus, the states with higher occupation numbers are suppressed in the athermal baths when compared to that of the thermal baths. Thus,  the average occupation number decreases for such a Kerr-athermal bath when compared with the thermal case, i.e., $\bar{n}_\beta(\nu,\zeta) < \bar{n}_\beta(\nu,\infty)$. This suppression gets stronger as $\zeta$ decreases. 
\\ \\
Now, for a positive transition energy $\mathcal{E}=\hbar\nu$, the ratio of relaxation to excitation rates is
\begin{align*}
    r_\alpha(\mathcal{E}) = \frac{\gamma_{+\nu,\alpha}}{\gamma_{-\nu,\alpha}} = \frac{\bar{n}_\beta(\nu, \zeta) + 1}{\bar{n}_\beta({\nu, \zeta})} = e^{\beta_{\text{eff},\alpha}\mathcal{E}},
\end{align*}
where $\beta_{\text{eff},\alpha} =[k_B T_{\text{eff},\alpha}]^{-1}$ defines the effective temperature associated with that transition.
As, there is a decrease in $\bar{n}_\beta(\nu, \zeta)$ with decrease in $\zeta$, we observe an increase in $r_\alpha(\mathcal{E})$. Consequently, the there is a increase in $\beta_{\text{eff},\alpha}$ and a decrease in the effective temperature $T_{\text{eff},\alpha}$. 
Thus, stronger Kerr-type preparation favors relaxation relative to excitation at each transition. These modified rate ratios provide physical intuition for the lower steady-state cold-qubit temperatures observed in our simulations compared with the thermal reference. The actual steady-state temperature is determined by the combined action of all three reservoirs and the inter-qubit interaction.}

\SuM{The additional resource relative to the thermal reference is the athermal preparation of the reservoir occupation distributions. Within the weak-coupling description with linear system–bath coupling used here, the dissipative rates depend on the mean occupations and spectral densities at the relevant transition frequencies. Other stationary, uncorrelated, number-diagonal reservoirs with the same mean occupations and spectral densities at these frequencies would therefore produce the same dissipative rates. Kerr-type preparation specifies the particular family of occupation distributions investigated in this work.}

\subsection{Scaling of  advantage in steady-state cooling with anharmonicity}
\label{subsec:scaling_T}

In the preceding sections, we have illustrated that \SuMo{anharmonicity in the athermal states} of the baths attached to the three qubits in the quantum refrigerator studied in this work, causes better steady-state cooling of the target qubit over the same with thermal baths, as depicted in Fig.~\ref{fig:tc-ssc}.
In this section, we explore how this advantage scales with the anharmonicity, 
and how fast it diminishes as we tend to the thermal limit. We denote the difference between the final steady-state temperature of the cold qubit and its initial temperature by \(\Delta\theta_c\). 
For this analysis, we confine ourselves to the parameter space where we can achieve steady-state cooling using thermal baths as well. 
The relevant parameters are specified in Sec.~\ref{subsec:ssc}. 
Within this regime, the considered quantum refrigerator always provides steady-state cooling irrespective of the strength of anharmonicity in the bath modes. 
From Fig.~\ref{fig:scaling_dT} we observe that for very small values of \(\zeta\) 
this advantage decays rapidly fast with decreasing anharmonicity. 
 In the range of anharmonicity 
 \( \zeta > 100 \), it decays very slowly, still providing a substantial steady-state cooling advantage over the \SuMo{thermal} case. 
This advantage of larger \(\Delta\theta_c\) in Fig.~\ref{fig:scaling_dT} asymptotically approaches \SuM{cooling attained with the thermal reference} in the limit of \(\zeta\rightarrow\infty\), i.e., when we revert back to the \SuMo{thermal} case.  
It is noticeable that even a very small amount of anharmonicity in the modes \SuMo{of the preparation Hamiltonians $H_{\text{prep},\alpha}$} is enough to aid better refrigeration of the target qubit, thereby establishing the superiority of using baths composed of \SuMo{athermal baths with anharmonicity in $H_{\text{prep},\alpha}$} for the purpose of quantum refrigeration compared to \SuMo{thermal} baths.

\subsection{Minimum attainable temperature}  
\label{subsec:polar}

\textcolor{black}{In this section, we investigate the capacity of a three-qubit refrigerator 
in terms of the minimum attainable temperature. For this purpose, one can exploit the potential of certain refrigeration scenarios, called transient cooling, where it is possible to achieve a much better cooling of the target qubit in the transient regime as compared to that in the steady state.
For this study, we work with the parameters as used in subsection~\ref{subsec:ssc}, which aid us in cooling at all times even when we incorporate \SuMo{thermal} baths in the scheme of quantum refrigeration. 
As discussed briefly in the aforementioned section, and observed in Fig.~\ref{fig:tc-ssc}(b), it is clear that for small anharmonicity (\(\zeta \gtrsim 50\)) the lowest temperature of the target qubit is attainable in the transient regime and one can attain the same minimum temperature in transient time using \SuMo{thermal} baths. 
For larger anharmonicity, the steady-state temperature of the cold qubit goes down past this transient temperature. 
\SuMo{Stronger anharmonicity of the preparation Hamiltonian $H_{\text{prep},\alpha}$ of the bath}, therefore, promises better cooling, and that too in the steady state. 
We denote the minimum attainable dimensionless temperature of the cold qubit by \(\theta^m\) and describe its variation with the inverse anharmonicity factor \(\zeta\) in Fig~\ref{fig:polar-temp}.
In this figure, we observe that for small values of \(\zeta\), the minimum temperature of the cold qubit increases with decreasing anharmonicity \SuMo{of $H_{\text{prep},\alpha}$}, and at \(\zeta\approx 50\), it reaches the minimum temperature of the cold qubit attained using harmonic baths. 
In the regime, $\zeta \gtrsim 50$, the target qubit attains a lower transient temperature before rising to its steady-state value. 
However, the same minimal transient temperature can also be achieved using \SuMo{thermal} baths, though it leads to higher temperatures in the steady state. \textcolor{black}{Note that in the variation of minimum attainable temperature $\theta^m$ with the anharmonicity of the baths $\zeta$, 
the transition value $\zeta \approx 50$ obtained here is not universal. This transition point depends on the chosen system-bath parameters. For instance, when $\omega_{w} = \omega_{c} = 2$ and all other parameters are kept the same as in Fig.~\ref{fig:polar-temp}, the transition occurs at $\zeta \approx 82$.}
Hence, we can conclude that the use of \SuMo{Kerr-athermal} baths for quantum refrigeration provides us with prospects of superior steady-state as well as transient cooling.}


\begin{figure*} 
\includegraphics[width=\textwidth]{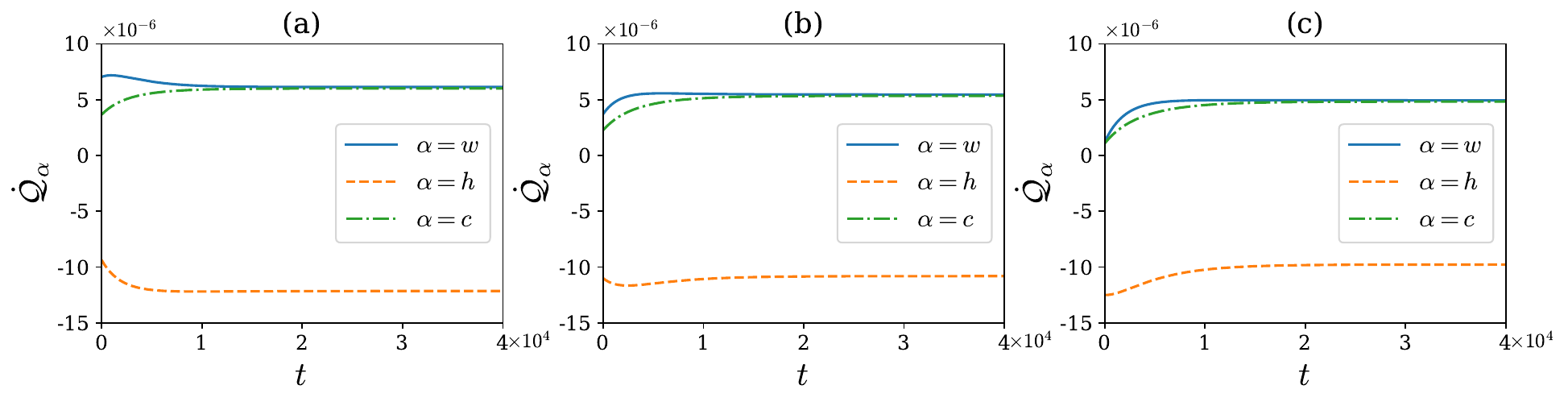}
\caption{Time-variation of the heat currents originating from the three baths constituting the quantum refrigerator. Time is shown along the horizontal axis and vertical axes shows the heats currents, \( \dot{ \mathcal{Q}}_{\alpha}\) (Eq.~\eqref{eq:heat-current}), for the three qubits distinguished by different \(\alpha\) as described in the legends of the plots.  Panel (a) depicts the relevant heat currents when we choose to use \SuMo{thermal} baths locally connected to the qubits. Panels (b) and (c) describe the same, but for the case of \SuMo{Kerr-athermal} baths with \(\zeta=100\) and \(\zeta=50\), respectively. As the values \(\zeta\) decrease from left to right in this figure, the bath-state move further away from the thermal regime. All the other relevant parameters are described in Sec.~\ref{subsec:ssc}. The vertical axes are in units of $K^2/\hbar$, while the horizontal axes are dimensionless.}
\label{fig:qchw}
\end{figure*}

\subsection{Heat currents and efficiencies of refrigeration}
\label{subsec:cop}

\begin{figure}
\includegraphics[width=0.5\textwidth]{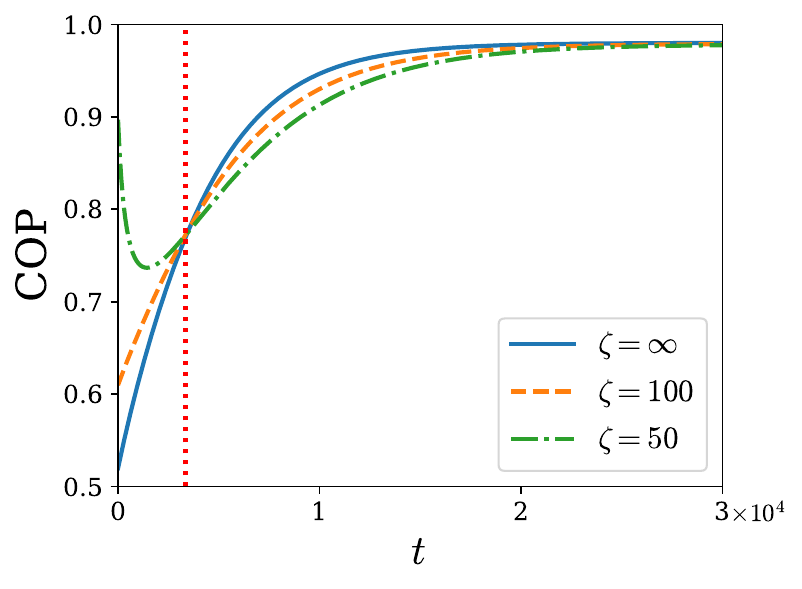}
\caption{Time variation of the coefficient of performance (COP) of the three-qubit quantum refrigerator with different anharmonicities in the \SuMo{preparation Hamiltonian $H_{\text{prep},\alpha}$ of the Kerr-athermal baths}. The horizontal axis shows time and the vertical axis presents COP for values of \(\zeta\) described in the legend. \(\zeta=\infty\) marks the harmonic limit. 
The red dotted vertical line signifies the time at which the COPs for different values of \(\zeta\) meet before they merge again in the steady state. The relevant system parameters are as described in Sec.~\ref{subsec:ssc}. All the quantities plotted along the horizontal and vertical axes are dimensionless.} 
\label{fig:cop} 
\end{figure}

Now we bring our attention to the thermodynamic quantities signifying the power and efficiency of the three-qubit quantum refrigerator studied in this paper. 
For this purpose, we first compute the heat currents of all three subsystems, which quantify the rate of heat flow from the local heat baths to the corresponding qubits.
The instantaneous heat currents are given by 
\begin{equation}
 \dot{ \mathcal{Q}}_{\alpha}(t) = \text{Tr}[H_{\text{ref}} \mathcal{D_{\alpha}}(\rho_s(t))],
\label{eq:heat-current}
\end{equation} 
where the dissipator corresponding to the qubit \(\alpha\) is given by \(\mathcal{D_{\alpha}}(\rho_s(t))\), as expressed in Eq.~\eqref{eq:D} and \(H_{\text{ref}}\) is the Hamiltonian of the refrigerator given in Eq.~\eqref{eq:ref}. 
The cooling power of the refrigerator is expressed by the heat current of the cold qubit at steady state: \( \dot{ \mathcal{Q}}_c(t\rightarrow\infty)\). 
A positive value of heat current indicates heat flowing out of the corresponding bath and into the interacting refrigerator. 


For the model of absorption refrigerator described in Sec.~\ref{sec:refg}, the coefficient of performance (COP) is measured by the ratio of the heat extracted from the cold qubit to the heat flow out of the work or engine qubit, i.e., the machine qubit. 
Therefore, COP of this quantum refrigerator is given by \(COP=\dot{ \mathcal{Q}}_c/\dot{ \mathcal{Q}}_w\). 
In Fig.~\ref{fig:qchw} we exhibit the cooling power (marked by \(\alpha=c\)) along with the heat currents from the hot and work qubits as well. 
From all the panels of Fig.~\ref{fig:qchw} we can see that the heat currents from the cold and work baths are positive and that of the hot bath is negative. 
This implies heat flow out of the cold and work bath and flow of heat into 
the hot bath
thereby cooling the target qubit as expected. 
However comparing Figs.~\ref{fig:qchw}(a) (harmonic-oscillator baths),~\ref{fig:qchw}(b) (\(\zeta=100\)), and~\ref{fig:qchw}(c) (\(\zeta=50\)), we observe that with increasing anharmonicity in the bath modes, the transient heat currents from the cold and work baths start with a smaller value, and the hot bath gets dumped with a higher heat current at short times. 
However, in steady state, all the heat currents achieve similar values irrespective of the anharmonicity of the \SuMo{athermal state of the baths}, and the total heat current from the whole system vanishes. 
This is reflected in the nature of the COP as well, demonstrated in Fig.~\ref{fig:cop}. 
We can see that the COP of the quantum refrigerator starts off with a substantially higher value for higher anharmonicity in \SuMo{$H_{\text{prep},\alpha}$, compared to the case of thermal baths being used in the three-qubit quantum refrigerator}. But as time goes on, the curves exhibiting the COP of the refrigerators meet at a point (marked by the red dotted line in Fig.~\ref{fig:cop}), followed by a higher COP in the case of the \SuMo{thermal baths}, until all the curves stabilize to the same COP in steady state. 

\SuM{Thus, the steady-state cold-qubit temperature is the principal figure of merit for the advantage demonstrated here. The engineered reservoir occupations modify the stationary populations, allowing the cold qubit to reach a lower temperature without an appreciable improvement in the steady-state COP. The attainable temperature, cooling power, and COP therefore characterize different aspects of the device's performance. The reported COP describes operation with the specified reservoirs and excludes the costs of preparing or maintaining their athermal states.}
\SuM{
\subsection{Realization of Kerr-athermal baths}
{Engineered reservoirs have been explored across a wide range of quantum thermodynamic settings.} Experimental control of Kerr nonlinearity in superconducting microwave cavities has enabled quantum-state collapse and revival~\cite{Kirchmair2013}, state preparation using tunable interactions~\cite{He2023}, and independent tuning of resonator frequency and Kerr strength~\cite{Miano2022}. Kerr-assisted cooling of mechanical oscillators has also been demonstrated experimentally~\cite{Bothner2022,Zoepfl2023} and investigated theoretically~\cite{DiazNaufal2025}. 
}

\SuM{
{Structured environments} provide several routes to modifying thermal-machine performance. Squeezed reservoirs can enhance absorption refrigeration~\cite{Correa2014, Klaers2017}, while reservoirs at effective negative temperatures~\cite{deAssis2019} have been investigated in heat engines. Quantum machines powered by correlated reservoirs have also been studied~\cite{DeChiara2020}. In our model, the reservoir states are number-diagonal, have decreasing occupation probabilities, and yield positive effective temperatures at the relevant transitions. The mechanism therefore requires neither squeezing nor population inversion. Our results provide a particular example of reservoir occupation engineering, rather than establishing a temperature advantage exclusive to Kerr-type preparation.
}

\SuM{For our model, a possible implementation would {require the preparation of} the required number-state populations and subsequently suppress the Kerr interaction to allow approximately harmonic evolution. Oscillator-state synthesis in superconducting resonators~\cite{Hofheinz2009} and tunable Kerr circuits~\cite{He2023} provide relevant experimental capabilities. Moreover, the preparation need not cover the entire reservoir spectrum. For the uncorrelated, number-diagonal baths and linear coupling considered here, the weak-coupling dissipative rates depend on the mean occupations and spectral densities at the allowed transition frequencies of the interacting refrigerator
~\cite{Alicki2015}. 
Engineering suitable frequency bands around these transitions can therefore reproduce the required rates while leaving the remaining modes thermal, provided that the reservoir correlation time remains short compared with the refrigerator relaxation time.
In current experiments like Ref.~\cite{He2023}, positive Kerr nonlinearities of around $5~\text{MHz}$ in resonators operating at approximately $4$-$5$~$\text{GHz}$ provide an illustrative scale of $\zeta\sim10^3$. Tuning anharmonicities have been demonstrated in superconducting unimon circuits~\cite{Hyyppa2022}, although applying our Kerr description would require checking the spectrum across the populated levels. 
{These examples collectively demonstrate that while reservoir engineering is experimentally accessible in various platforms, preparing and maintaining the specific occupation structure considered here represents a challenge that ongoing experimental advances are well positioned to address.}
}

\section{Conclusion}
\label{conclusion}

In this work, we have demonstrated the functional advantage of using Kerr-\SuMo{athermal} baths in the scheme of three-qubit quantum refrigerators. Significantly better steady-state cooling of the target qubit is obtained in this case, compared to that in the case of \SuMo{thermal} baths. 
We have confined our study to the limit where system-bath coupling remains in the Markovian regime, and hence the GKSL master equation can faithfully capture the dynamics of the system.

\SuM{We have studied a three-qubit cooling device coupled to harmonic baths prepared in stationary Kerr-athermal states. For the parameters investigated, these baths lower the steady-state temperature of the target qubit compared with the thermal-bath case. In the transient-cooling regime, sufficiently strong Kerr-type preparation lowers the steady-state temperature below the initial value, although the early transient minimum is not improved in the examples considered. In the steady-state cooling regime, the steady-state temperature can become lower than the early transient minimum of the thermal reference. The main benefit therefore lies in the temperature attained in the steady state, while the early transient minimum shows little or no improvement. This reduction in the steady-state temperature does not imply higher thermodynamic efficiency. In the examples observed in steady-state cooling regime, the cooling power decreases while the steady-state coefficient of performance remains nearly unchanged compared with the thermal-bath case. The temperature advantage relative to the chosen thermal reference is attributed to the engineered reservoir occupations.}
We have shown that even athermality due to a very small amount of anharmonicity in the bath modes of the preparation Hamiltonian can help us acquire a reasonable advantage over using \SuMo{thermal} baths, in steady-state regime. 
We also studied how this advantage scales with the anharmonicity. 
%
{\color{black}
Nonlinear environments have been extensively studied, both theoretically and experimentally, across a wide range of quantum platforms~\cite{PhysRevLett.79.1467,Winkler2006,Firstenberg2013, Leghtas2015, PhysRevLett.124.213601}. In this context, it is natural and experimentally relevant to consider the implementation of quantum refrigerators coupled to Kerr-athermal environments. 
We further believe that this study makes it plausible that anharmonicity in bath states will prove useful in the investigation of other quantum thermal machines, such as quantum heat pumps.

}

\section*{Acknowledgements}
 We acknowledge computations performed using Armadillo~\cite{Sanderson2016, Sanderson1}.
The research of A.G. was supported by the INFOSYS scholarship and the Alexander von Humboldt Foundation.
D.R. acknowledges support from Science and Engineering Research Board (SERB), Department of Science and Technology (DST), under the sanction No. SRG/2021/002316-G.
We acknowledge partial support from the Department of Science and Technology, Government of India through the QuEST grant (grant number DST/ICPS/QUST/Theme-3/2019/120). 

\appendix
\section*{Appendices}
\textcolor{black}{\section{Explicit expressions of the dynamical equations}
\label{appen:B}}
\textcolor{black}{The explicit forms of the Lindblad operators defined in Eq.~(\ref{eq:A-op}) of the main text are given by 
\begin{align*}
    A_{\omega_c,c} &= |1\rangle\langle 4| + |5\rangle\langle 3|, \\
    A_{\omega_c+g,c} &= (|2\rangle\langle 8| - |7\rangle\langle 6|)/\sqrt2, \\
    A_{\omega_c-g,c} &= (|2\rangle\langle 7| + |8\rangle\langle 6|)/\sqrt2, \\
    A_{\omega_h,h} &= |2\rangle\langle 5| + |4\rangle\langle 6|, \\
    A_{\omega_h+g,h} &= (|7\rangle\langle 3| + |1\rangle\langle 8|)/\sqrt2, \\
    A_{\omega_h-g,h} &= (|8\rangle\langle 3| - |1\rangle\langle 7|)/\sqrt2, \\
    A_{\omega_w,w} &= |1\rangle\langle 2| + |6\rangle\langle 3|, \\
    A_{\omega_w+g,w} &= (|4\rangle\langle 8| - |7\rangle\langle 5|)/\sqrt2, \\
    A_{\omega_w-g,w} &= (|4\rangle\langle 7| + |8\rangle\langle 5|)/\sqrt2. \\
\end{align*}
Here, 
$|1\rangle = |0_c0_h0_w\rangle$ ,
    $ |2\rangle = |0_c0_h1_w\rangle$,  
    $ |3\rangle = |1_c1_h1_w\rangle$,
    $ |4\rangle = |1_c0_h0_w\rangle$,
    $ |5\rangle = |0_c1_h1_w\rangle$,
    $ |6\rangle = |1_c1_h0_w\rangle$, 
    $ |7\rangle = (|0_c1_h0_w\rangle - |1_c0_h1_w\rangle)/\sqrt2$, and
    $ |8\rangle = (|0_c1_h0_w\rangle + |1_c0_h1_w\rangle)/\sqrt2$ 
     are the eigenstates of the Hamiltonian $H_\text{ref}$. }
     
\textcolor{black}{In order to find the steady-state solution, we need to solve $\rho_{ss}$ for
\begin{align}
\frac{d\rho_{ss}}{dt} = - \frac{i}{K}[H_{\text{ref}},\rho_{ss}] +  \frac{\hbar}{K}\sum_{\alpha = c,h,w} \mathcal{D}_\alpha(\rho_{ss}) = 0.    
\end{align}
 The steady-state solutions are the right eigenvector of the $\mathcal{L}$ operator with eigenvalues such $\lambda_{ss}$, such that $\text{Re}[{\lambda_{ss}}] = 0$. We find that $\rho_{ss}$ is not oscillatory in nature, thus $\text{Im}[\lambda_{ss}] = 0$. Thus, we have $[H_{\text{ref}},\rho_{ss}] = 0$ and the state is of the form
\begin{align*}
    \rho_{ss} = \sum_i \rho_i |i\rangle\langle i|,
\end{align*}
where $\{|i\rangle\}$ are the eigenstates of the Hamiltonian $H_\text{ref}$. }
\textcolor{black}{Hence, we have 
\begin{align*}
   \mathcal{D}_c(\rho_{ss}) + \mathcal{D}_h(\rho_{ss}) + \mathcal{D}_w(\rho_{ss}) = 0.
\end{align*}
From this we get eight linear equations : 
\begin{strip}
\begin{align}
\label{ss-eq1}
   &-D_1\rho_1 + \ga\rho_2 + \gm \rho_4 + (\gz/2) \rho_7 + (\gy/2)\rho_8  = 0, \nonumber \\
  & \gan \rho_1 - D_2 \rho_2 + \gx \rho_5  + (\go/2)\rho_7+ (\gn/2)\rho_8  = 0, \nonumber \\
   & -D_3\rho_3  + \gmn \rho_5 + \gan \rho_6 + (\gyn/2)\rho_7 + (\gzn/2)\rho_8  = 0, \nonumber \\
  &  \gmn \rho_1 - D_4\rho_4 + \gx\rho_6 + (\gc/2)\rho_7 + (\gb/2)\rho_8 = 0, \nonumber \\
  &   \gxn \rho_2 + \gm\rho_3 - D_5\rho_5 + (\gbn/2)\rho_7 + (\gcn/2)\rho_8 = 0, \nonumber \\
 &   \ga \rho_3 + \gxn\rho_4 -D_6\rho_6 + (\gnn/2)\rho_7 + (\gon/2)\rho_8 = 0, \nonumber \\
 &   (\gzn/2)\rho_1 + (\gon/2)\rho_2 + (\gy/2)\rho_3 + (\gcn/2)\rho_4 + (\gb/2)\rho_5 + (\gn/2)\rho_6 -D_7\rho_7  = 0, \nonumber \\
 &   (\gyn/2)\rho_1 +(\gnn/2)\rho_2 +(\gz/2)\rho_3 +(\gbn/2)\rho_4 +(\gc/2)\rho_5 +(\go/2)\rho_6  - D_8\rho_8  = 0, 
\end{align}
with
\begin{align*}
&D_1 = \gan + \gyn/2 + \gzn/2 + \gmn,\\
&D_2 = \ga + \gxn + \gnn/2 + \gon/2,\\
&D_3 = \ga + \gy/2 + \gz/2 + \gm, \\
&D_4 = \gbn/2 + \gcn/2 + \gxn + \gm,\\
&D_5 = \gb/2 + \gc/2 + \gx + \gmn,\\
&D_6 = \gan + \gx + \gn/2 + \go/2,\\
&D_7 = (\gbn + \gc + \gyn + \gz +\gnn + \go)/2,\\
&D_8= (\gb + \gcn + \gy +  \gzn +\gn + \gon)/2.
\end{align*}
\end{strip}
In addition to these, we have the trace condition, 
\begin{align}
\label{ss-eq2}
    \rho_1 + \rho_2 + \rho_3 + \rho_4 + \rho_5 + \rho_6 + \rho_7 + \rho_8  = 1.
\end{align}
Solving both Eqs. \eqref{ss-eq1} and \eqref{ss-eq2} analytically is a bit challenging. We numerically solve them and present the steady-state temperature of the cold qubit here.}

\section{Average particle number of Kerr-\SuMo{athermal} baths}
\label{appen:A}
The average number of particles  for \SuMo{Kerr-athermal baths} is given by
\begin{align}
    \label{eq:n-w}
     \hspace{-10mm}&\braket{n_{\beta}(\mathcal{E})} = \braket{b_{\mathcal{E}}^{\dagger}b_{\mathcal{E}}} = \ddfrac{\text{Tr}[b_{\mathcal{E}}^{\dagger} b_{\mathcal{E}} \exp(-\beta H_{\text{prep}})]}{\text{Tr}[\exp(-\beta H_{\text{prep}})]}  \nonumber \\
     \hspace{-10mm}&= \frac{\sum\limits_{\{n_k\}} \braket{\{n_k\}|\hat n_{\mathcal{E}}\exp\Big[-\hslash\beta\sum\limits_k \left(\Omega_k \hat n_k+ \frac{\Omega_k}{\zeta}  (\hat n_k)^2 \right)\Big]|\{n_k\}}}{\sum\limits_{\{n_k\}} \braket{\{n_k\}|\exp\Big[-\hslash\beta\sum\limits_k \left(\Omega_k \hat n_k + \frac{\Omega_k}{\zeta}  (\hat n_k)^2 \right)\Big]|\{n_k\}}},
\end{align}
where \(\{n_k\}=\{n_{1,k},n_{2,k},n_{3,k},\ldots\}\) denotes the collection of occupation numbers specifying the oscillator states. Since the oscillators are bosonic, each occupation number can take any non-negative integer value. Also, we have $\hat n_{k} = b_k^\dagger b_k$.
Since \([b_k^{\dagger}b_k, b_{\mathcal{E}}]=[b_k^{\dagger}b_k, b_{\mathcal{E}}^{\dagger}] = 0 \text{ for all } k \neq \mathcal{E}\), the exponent in Eq.~\eqref{eq:n-w} factorizes, and it can be written as 
\begin{align*}
    &  \exp\Big[-\hslash\beta\sum_k \left(\Omega_k b_k^{\dagger}b_k + \frac{\Omega_k}{\zeta}  (b_k^{\dagger}b_k)^2 \right)\Big] \nonumber \\
    & = \prod_k \exp\Big[-\hslash\beta\left(\Omega_k b_k^{\dagger}b_k + \frac{\Omega_k}{\zeta}  (b_k^{\dagger}b_k)^2 \right)\Big].
\end{align*}
Clearly the terms in the  numerator and denominator for \(k \neq \mathcal{E}\) in Eq.~\eqref{eq:n-w} are the same and hence cancel each other, so that we are only left with the terms with frequency \(\mathcal{E}\) to sum over. Clearly, it follows that, 
\begin{align*}
\label{eq:nw}
    \hspace{-10mm}& \braket{n_{\beta}(\mathcal{E})}  \nonumber\\
    \hspace{-10mm}&= \ddfrac{\sum\limits_{n_{\mathcal{E}}=0}^{\infty}\bra{n_{\mathcal{E}}}b_{\mathcal{E}}^{\dagger}b_{\mathcal{E}}\exp{\Big[-\hslash\beta\left(\mathcal{E} b_{\mathcal{E}}^{\dagger}b_{\mathcal{E}}  + \frac{\mathcal{E}}{\zeta}  (b_{\mathcal{E}}^{\dagger}b_{\mathcal{E}})^2 \right)\Big]}\ket{n_{\mathcal{E}}}}{\sum\limits_{n_{\mathcal{E}}=0}^{\infty}\bra{n_{\mathcal{E}}}\exp\Big[-\hslash\beta\left(\mathcal{E} b_{\mathcal{E}}^{\dagger}b_{\mathcal{E}}  + \frac{\mathcal{E}}{\zeta}  (b_{\mathcal{E}}^{\dagger}b_{\mathcal{E}})^2 \right)\Big]\ket{n_{\mathcal{E}}}}  \nonumber\\ 
    \hspace{-10mm}& =  \ddfrac{\sum\limits_{n_{\mathcal{E}}=0}^{\infty}n_{\mathcal{E}}\exp\Big[-\hslash\beta\left(\mathcal{E} n_{\mathcal{E}}  + \frac{\mathcal{E}}{\zeta}  n_{\mathcal{E}}^2 \right)\Big]}{\sum\limits_{n_{\mathcal{E}}=0}^{\infty}\exp\Big[-\hslash\beta\left(\mathcal{E} n_{\mathcal{E}}  + \frac{\mathcal{E}}{\zeta}  n_{\mathcal{E}}^2 \right)\Big]}.\hfill
\end{align*}

\printcredits

\printbibliography

@article{Brask2015,
  title = {Small quantum absorption refrigerator in the transient regime: Time scales, enhanced cooling, and entanglement},
  author = {Brask, Jonatan Bohr and Brunner, Nicolas},
  journal = {Phys. Rev. E},
  volume = {92},
  issue = {6},
  pages = {062101},
  numpages = {7},
  year = {2015},
  month = {Dec},
  publisher = {American Physical Society},
  doi = {10.1103/PhysRevE.92.062101},
  url = {https://link.aps.org/doi/10.1103/PhysRevE.92.062101}
}

@article{Brunner2012,
  title = {Virtual qubits, virtual temperatures, and the foundations of thermodynamics},
  author = {Brunner, Nicolas and Linden, Noah and Popescu, Sandu and Skrzypczyk, Paul},
  journal = {Phys. Rev. E},
  volume = {85},
  issue = {5},
  pages = {051117},
  numpages = {14},
  year = {2012},
  month = {May},
  publisher = {American Physical Society},
  doi = {10.1103/PhysRevE.85.051117},
  url = {https://link.aps.org/doi/10.1103/PhysRevE.85.051117}
}

@article{Brunner2014,
  title = {Entanglement enhances cooling in microscopic quantum refrigerators},
  author = {Brunner, Nicolas and Huber, Marcus and Linden, Noah and Popescu, Sandu and Silva, Ralph and Skrzypczyk, Paul},
  journal = {Phys. Rev. E},
  volume = {89},
  issue = {3},
  pages = {032115},
  numpages = {6},
  year = {2014},
  month = {Mar},
  publisher = {American Physical Society},
  doi = {10.1103/PhysRevE.89.032115},
  url = {https://link.aps.org/doi/10.1103/PhysRevE.89.032115}
}

@article{Chen2012,
doi = {10.1209/0295-5075/97/40003},
url = {https://dx.doi.org/10.1209/0295-5075/97/40003},
year = {2012},
month = {feb},
publisher = {},
volume = {97},
number = {4},
pages = {40003},
author = {Yi-Xin Chen and Sheng-Wen Li},
title = {Quantum refrigerator driven by current noise},
journal = {Europhysics Letters}
}

@article{Correa2013,
  title = {Performance bound for quantum absorption refrigerators},
  author = {Correa, Luis A. and Palao, Jos\'e P. and Adesso, Gerardo and Alonso, Daniel},
  journal = {Phys. Rev. E},
  volume = {87},
  issue = {4},
  pages = {042131},
  numpages = {9},
  year = {2013},
  month = {Apr},
  publisher = {American Physical Society},
  doi = {10.1103/PhysRevE.87.042131},
  url = {https://link.aps.org/doi/10.1103/PhysRevE.87.042131}
}

@Article{Correa2014,
author={Correa, Luis A. and Palao, Jos{\'e} P. and Alonso, Daniel and Adesso, Gerardo},
title={Quantum-enhanced absorption refrigerators},
journal={Scientific Reports},
year={2014},
month={Feb},
day={04},
volume={4},
number={1},
pages={3949},
doi={10.1038/srep03949},
url={https://doi.org/10.1038/srep03949}
}

@article{Das2019,
doi = {10.1209/0295-5075/125/20007},
url = {https://dx.doi.org/10.1209/0295-5075/125/20007},
year = {2019},
month = {feb},
publisher = {EDP Sciences, IOP Publishing and Società Italiana di Fisica},
volume = {125},
number = {2},
pages = {20007},
author = {Sreetama Das and Avijit Misra and Amit Kumar Pal and Aditi Sen(De) and Ujjwal Sen},
title = {Necessarily transient quantum refrigerator},
journal = {Europhysics Letters}
}

@article{Felce2020,
  title = {Quantum Refrigeration with Indefinite Causal Order},
  author = {Felce, David and Vedral, Vlatko},
  journal = {Phys. Rev. Lett.},
  volume = {125},
  issue = {7},
  pages = {070603},
  numpages = {6},
  year = {2020},
  month = {Aug},
  publisher = {American Physical Society},
  doi = {10.1103/PhysRevLett.125.070603},
  url = {https://link.aps.org/doi/10.1103/PhysRevLett.125.070603}
}

@article{Goold2016,
doi = {10.1088/1751-8113/49/14/143001},
url = {https://dx.doi.org/10.1088/1751-8113/49/14/143001},
year = {2016},
month = {feb},
publisher = {IOP Publishing},
volume = {49},
number = {14},
pages = {143001},
author = {John Goold and Marcus Huber and Arnau Riera and Lídia del Rio and Paul Skrzypczyk},
title = {The role of quantum information in thermodynamics—a topical review},
journal = {Journal of Physics A: Mathematical and Theoretical}
}

@article{Giazotto2006,
  title = {Opportunities for mesoscopics in thermometry and refrigeration: Physics and applications},
  author = {Giazotto, Francesco and Heikkil\"a, Tero T. and Luukanen, Arttu and Savin, Alexander M. and Pekola, Jukka P.},
  journal = {Rev. Mod. Phys.},
  volume = {78},
  issue = {1},
  pages = {217--274},
  numpages = {0},
  year = {2006},
  month = {Mar},
  publisher = {American Physical Society},
  doi = {10.1103/RevModPhys.78.217},
  url = {https://link.aps.org/doi/10.1103/RevModPhys.78.217}
}

@article{Hewgill2020,
  title = {Three-qubit refrigerator with two-body interactions},
  author = {Hewgill, Adam and Gonz\'alez, J. Onam and Palao, Jos\'e P. and Alonso, Daniel and Ferraro, Alessandro and De Chiara, Gabriele},
  journal = {Phys. Rev. E},
  volume = {101},
  issue = {1},
  pages = {012109},
  numpages = {10},
  year = {2020},
  month = {Jan},
  publisher = {American Physical Society},
  doi = {10.1103/PhysRevE.101.012109},
  url = {https://link.aps.org/doi/10.1103/PhysRevE.101.012109}
}

@article{Hofer2016,
  title = {Autonomous quantum refrigerator in a circuit QED architecture based on a Josephson junction},
  author = {Hofer, Patrick P. and Perarnau-Llobet, Mart\'{\i} and Brask, Jonatan Bohr and Silva, Ralph and Huber, Marcus and Brunner, Nicolas},
  journal = {Phys. Rev. B},
  volume = {94},
  issue = {23},
  pages = {235420},
  numpages = {7},
  year = {2016},
  month = {Dec},
  publisher = {American Physical Society},
  doi = {10.1103/PhysRevB.94.235420},
  url = {https://link.aps.org/doi/10.1103/PhysRevB.94.235420}
}

@article{Hofer2016heateng,
  title = {Quantum heat engine based on photon-assisted Cooper pair tunneling},
  author = {Hofer, Patrick P. and Souquet, J.-R. and Clerk, A. A.},
  journal = {Phys. Rev. B},
  volume = {93},
  issue = {4},
  pages = {041418},
  numpages = {5},
  year = {2016},
  month = {Jan},
  publisher = {American Physical Society},
  doi = {10.1103/PhysRevB.93.041418},
  url = {https://link.aps.org/doi/10.1103/PhysRevB.93.041418}
}

@Article{Kosloff2013,
AUTHOR = {Kosloff, Ronnie},
TITLE = {Quantum Thermodynamics: A Dynamical Viewpoint},
JOURNAL = {Entropy},
VOLUME = {15},
YEAR = {2013},
NUMBER = {6},
PAGES = {2100--2128},
URL = {https://www.mdpi.com/1099-4300/15/6/2100},
ISSN = {1099-4300},
DOI = {10.3390/e15062100}
}

@article{Kosloff2014,
author = {Kosloff, Ronnie and Levy, Amikam},
title = {Quantum Heat Engines and Refrigerators: Continuous Devices},
journal = {Annual Review of Physical Chemistry},
volume = {65},
number = {1},
pages = {365-393},
year = {2014},
doi = {10.1146/annurev-physchem-040513-103724},
    note ={PMID: 24689798},
URL = {https://doi.org/10.1146/annurev-physchem-040513-103724},
eprint = {https://doi.org/10.1146/annurev-physchem-040513-103724}
}

@article{Levy2012,
  title = {Quantum Absorption Refrigerator},
  author = {Levy, Amikam and Kosloff, Ronnie},
  journal = {Phys. Rev. Lett.},
  volume = {108},
  issue = {7},
  pages = {070604},
  numpages = {5},
  year = {2012},
  month = {Feb},
  publisher = {American Physical Society},
  doi = {10.1103/PhysRevLett.108.070604},
  url = {https://link.aps.org/doi/10.1103/PhysRevLett.108.070604}
}

@article{Linden2010,
  title = {How Small Can Thermal Machines Be? The Smallest Possible Refrigerator},
  author = {Linden, Noah and Popescu, Sandu and Skrzypczyk, Paul},
  journal = {Phys. Rev. Lett.},
  volume = {105},
  issue = {13},
  pages = {130401},
  numpages = {4},
  year = {2010},
  month = {Sep},
  publisher = {American Physical Society},
  doi = {10.1103/PhysRevLett.105.130401},
  url = {https://link.aps.org/doi/10.1103/PhysRevLett.105.130401}
}

@misc{Linden2010smallest,
      title={The smallest possible heat engines}, 
      author={Noah Linden and Sandu Popescu and Paul Skrzypczyk},
      year={2010},
      eprint={1010.6029},
      archivePrefix={arXiv},
      primaryClass={quant-ph}
}

@article{Mitchison2015,
doi = {10.1088/1367-2630/17/11/115013},
url = {https://dx.doi.org/10.1088/1367-2630/17/11/115013},
year = {2015},
month = {nov},
publisher = {IOP Publishing},
volume = {17},
number = {11},
pages = {115013},
author = {Mark T Mitchison and Mischa P Woods and Javier Prior and Marcus Huber},
title = {Coherence-assisted single-shot cooling by quantum absorption refrigerators},
journal = {New J. Phys.}
}

@article{Mitchison2016,
doi = {10.1088/2058-9565/1/1/015001},
url = {https://dx.doi.org/10.1088/2058-9565/1/1/015001},
year = {2016},
month = {aug},
publisher = {IOP Publishing},
volume = {1},
number = {1},
pages = {015001},
author = {Mark T Mitchison and Marcus Huber and Javier Prior and Mischa P Woods and Martin B Plenio},
title = {Realising a quantum absorption refrigerator with an atom-cavity system},
journal = {Quantum Science and Technology}
}

@Inbook{Mitchison2018,
author="Mitchison, Mark T.
and Potts, Patrick P.",
editor="Binder, Felix
and Correa, Luis A.
and Gogolin, Christian
and Anders, Janet
and Adesso, Gerardo",
title="Physical Implementations of Quantum Absorption Refrigerators",
bookTitle="Thermodynamics in the Quantum Regime: Fundamental Aspects and New Directions",
year="2018",
publisher="Springer International Publishing",
address="Cham",
pages="149--174",
isbn="978-3-319-99046-0",
doi="10.1007/978-3-319-99046-0_6",
url="https://doi.org/10.1007/978-3-319-99046-0_6"
}

@Book{Nielsen2011,
title = {Quantum Computation and Quantum Information: 10th Anniversary Edition},
author = {M. A. Nielsen and I. L. Chuang},
doi = {},
publisher = {Cambridge University Press},
number = {},
volume = {},
year = 2011,
month = {},
pages = {},
}

@article{Nie2022,
  title = {Experimental Realization of a Quantum Refrigerator Driven by Indefinite Causal Orders},
  author = {Nie, Xinfang and Zhu, Xuanran and Huang, Keyi and Tang, Kai and Long, Xinyue and Lin, Zidong and Tian, Yu and Qiu, Chudan and Xi, Cheng and Yang, Xiaodong and Li, Jun and Dong, Ying and Xin, Tao and Lu, Dawei},
  journal = {Phys. Rev. Lett.},
  volume = {129},
  issue = {10},
  pages = {100603},
  numpages = {6},
  year = {2022},
  month = {Sep},
  publisher = {American Physical Society},
  doi = {10.1103/PhysRevLett.129.100603},
  url = {https://link.aps.org/doi/10.1103/PhysRevLett.129.100603}
}

@book{Petruccione2002,
  address = {Great Clarendon Street},
  author = {Breuer, H. P. and Petruccione, F.},
  nota = {GK: excellent textbook!},
  publisher = {Oxford University Press},
  title = {The theory of open quantum systems},
  year = 2002
}

@book{Rivas2012,
  title={Open quantum systems},
  author={Rivas, Angel and Huelga, Susana F},
  volume={10},
  year = {2012},
  publisher={Springer}
}

@article{Skrzypczyk2011,
doi = {10.1088/1751-8113/44/49/492002},
url = {https://dx.doi.org/10.1088/1751-8113/44/49/492002},
year = {2011},
month = {nov},
publisher = {IOP Publishing},
volume = {44},
number = {49},
pages = {492002},
author = {Paul Skrzypczyk and Nicolas Brunner and Noah Linden and Sandu Popescu},
title = {The smallest refrigerators can reach maximal efficiency},
journal = {Journal of Physics A: Mathematical and Theoretical}
}

@article{Uzdin2015,
  title = {Equivalence of Quantum Heat Machines, and Quantum-Thermodynamic Signatures},
  author = {Uzdin, Raam and Levy, Amikam and Kosloff, Ronnie},
  journal = {Phys. Rev. X},
  volume = {5},
  issue = {3},
  pages = {031044},
  numpages = {21},
  year = {2015},
  month = {Sep},
  publisher = {American Physical Society},
  doi = {10.1103/PhysRevX.5.031044},
  url = {https://link.aps.org/doi/10.1103/PhysRevX.5.031044}
}

@article{Venturelli2013,
  doi = {10.1103/physrevlett.110.256801},
  url = {https://doi.org/10.1103/physrevlett.110.256801},
  year = {2013},
  month = jun,
  publisher = {American Physical Society ({APS})},
  volume = {110},
  number = {25},
  author = {Davide Venturelli and Rosario Fazio and Vittorio Giovannetti},
  title = {Minimal Self-Contained Quantum Refrigeration Machine Based on Four Quantum Dots},
  journal = {Physical Review Letters}
}

@article{Vinjanampathy2016,
author = {Sai Vinjanampathy and Janet Anders},
title = {Quantum thermodynamics},
journal = {Contemporary Physics},
volume = {57},
number = {4},
pages = {545-579},
year = {2016},
publisher = {Taylor & Francis},
doi = {10.1080/00107514.2016.1201896},
URL = {https://doi.org/10.1080/00107514.2016.1201896}
}

@article{Yu2014,
  doi = {10.1103/physreve.90.052142},
  url = {https://doi.org/10.1103/physreve.90.052142},
  year = {2014},
  month = nov,
  publisher = {American Physical Society ({APS})},
  volume = {90},
  number = {5},
  author = {Chang-shui Yu and Qing-yao Zhu},
  title = {Re-examining the self-contained quantum refrigerator in the strong-coupling regime},
  journal = {Physical Review E}
}

@article{diode,
  title = {Quantum thermal diode based on two interacting spinlike systems under different excitations},
  author = {Ordonez-Miranda, Jose and Ezzahri, Youn\`es and Joulain, Karl},
  journal = {Phys. Rev. E},
  volume = {95},
  issue = {2},
  pages = {022128},
  numpages = {7},
  year = {2017},
  month = {Feb},
  publisher = {American Physical Society},
  doi = {10.1103/PhysRevE.95.022128},
  url = {https://link.aps.org/doi/10.1103/PhysRevE.95.022128}
}

@article{refrigerator2,
  title = {Optimal refrigerator},
  author = {Allahverdyan, Armen E. and Hovhannisyan, Karen and Mahler, Guenter},
  journal = {Phys. Rev. E},
  volume = {81},
  issue = {5},
  pages = {051129},
  numpages = {12},
  year = {2010},
  month = {May},
  publisher = {American Physical Society},
  doi = {10.1103/PhysRevE.81.051129},
  url = {https://link.aps.org/doi/10.1103/PhysRevE.81.051129}
}

@article{Silva,
  title = {Small quantum absorption refrigerator with reversed couplings},
  author = {Silva, Ralph and Skrzypczyk, Paul and Brunner, Nicolas},
  journal = {Phys. Rev. E},
  volume = {92},
  issue = {1},
  pages = {012136},
  numpages = {6},
  year = {2015},
  month = {Jul},
  publisher = {American Physical Society},
  doi = {10.1103/PhysRevE.92.012136},
  url = {https://link.aps.org/doi/10.1103/PhysRevE.92.012136}
}

@article{refrigerator3,
  title = {Absorption refrigerators based on Coulomb-coupled single-electron systems},
  author = {Erdman, Paolo Andrea and Bhandari, Bibek and Fazio, Rosario and Pekola, Jukka P. and Taddei, Fabio},
  journal = {Phys. Rev. B},
  volume = {98},
  issue = {4},
  pages = {045433},
  numpages = {10},
  year = {2018},
  month = {Jul},
  publisher = {American Physical Society},
  doi = {10.1103/PhysRevB.98.045433},
  url = {https://link.aps.org/doi/10.1103/PhysRevB.98.045433}
}

@article{Avijit,
doi = {10.1088/2058-9565/ab8d89},
url = {https://dx.doi.org/10.1088/2058-9565/ab8d89},
year = {2020},
month = {may},
publisher = {IOP Publishing},
volume = {5},
number = {3},
pages = {035006},
author = {M Tahir Naseem and Avijit Misra and Özgür E Müstecaplıoğlu},
title = {Two-body quantum absorption refrigerators with optomechanical-like interactions},
journal = {Quantum Science and Technology}
}

@article{refrigerator4,
  title = {Smallest quantum thermal machine: The effect of strong coupling and distributed thermal tasks},
  author = {Man, Zhong-Xiao and Xia, Yun-Jie},
  journal = {Phys. Rev. E},
  volume = {96},
  issue = {1},
  pages = {012122},
  numpages = {9},
  year = {2017},
  month = {Jul},
  publisher = {American Physical Society},
  doi = {10.1103/PhysRevE.96.012122},
  url = {https://link.aps.org/doi/10.1103/PhysRevE.96.012122}
}

@article{refrigerator5,
  title = {Cooling condition for multilevel quantum absorption refrigerators},
  author = {Friedman, Hava Meira and Segal, Dvira},
  journal = {Phys. Rev. E},
  volume = {100},
  issue = {6},
  pages = {062112},
  numpages = {11},
  year = {2019},
  month = {Dec},
  publisher = {American Physical Society},
  doi = {10.1103/PhysRevE.100.062112},
  url = {https://link.aps.org/doi/10.1103/PhysRevE.100.062112}
}

@article{refrigerator6,
  title = {Enabling the self-contained refrigerator to work beyond its limits by filtering the reservoirs},
  author = {He, Zi-chen and Huang, Xin-yun and Yu, Chang-shui},
  journal = {Phys. Rev. E},
  volume = {96},
  issue = {5},
  pages = {052126},
  numpages = {10},
  year = {2017},
  month = {Nov},
  publisher = {American Physical Society},
  doi = {10.1103/PhysRevE.96.052126},
  url = {https://link.aps.org/doi/10.1103/PhysRevE.96.052126}
}

@article{refrigerator7,
doi = {10.1088/1367-2630/aac688},
url = {https://dx.doi.org/10.1088/1367-2630/aac688},
year = {2018},
month = {jun},
publisher = {IOP Publishing},
volume = {20},
number = {6},
pages = {063005},
author = {Jian-Ying Du and Fu-Lin Zhang},
title = {Nonequilibrium quantum absorption refrigerator},
journal = {New J. Phys.}
}

@article{qspeedlimit,
  title = {Quantum speed limit constraints on a nanoscale autonomous refrigerator},
  author = {Mukhopadhyay, Chiranjib and Misra, Avijit and Bhattacharya, Samyadeb and Pati, Arun Kumar},
  journal = {Phys. Rev. E},
  volume = {97},
  issue = {6},
  pages = {062116},
  numpages = {8},
  year = {2018},
  month = {Jun},
  publisher = {American Physical Society},
  doi = {10.1103/PhysRevE.97.062116},
  url = {https://link.aps.org/doi/10.1103/PhysRevE.97.062116}
}

@article{Scarani,
  title = {Refrigeration beyond weak internal coupling},
  author = {Seah, Stella and Nimmrichter, Stefan and Scarani, Valerio},
  journal = {Phys. Rev. E},
  volume = {98},
  issue = {1},
  pages = {012131},
  numpages = {7},
  year = {2018},
  month = {Jul},
  publisher = {American Physical Society},
  doi = {10.1103/PhysRevE.98.012131},
  url = {https://link.aps.org/doi/10.1103/PhysRevE.98.012131}
}

@article{article,
author = {Sanderson, C. and Curtin, R.},
year = {2016},
month = {07},
pages = {26},
title = {Armadillo: A template-based C++ library for linear algebra},
volume = {1},
journal = {Journal of Open Source Software},
doi = {10.21105/joss.00026}
}

@article{Yuan,
title = {A review on the electroluminescence properties of quantum-dot light-emitting diodes},
journal = {Organic Electronics},
volume = {90},
pages = {106086},
year = {2021},
issn = {1566-1199},
doi = {https://doi.org/10.1016/j.orgel.2021.106086},
url = {https://www.sciencedirect.com/science/article/pii/S1566119921000264},
author = {Qilin Yuan and Ting Wang and Panlong Yu and Hanzhuang Zhang and Han Zhang and Wenyu Ji}
}

@article{Joulain,
  title = {Quantum Thermal Transistor},
  author = {Joulain, Karl and Drevillon, J\'er\'emie and Ezzahri, Youn\`es and Ordonez-Miranda, Jose},
  journal = {Phys. Rev. Lett.},
  volume = {116},
  issue = {20},
  pages = {200601},
  numpages = {5},
  year = {2016},
  month = {May},
  publisher = {American Physical Society},
  doi = {10.1103/PhysRevLett.116.200601},
  url = {https://link.aps.org/doi/10.1103/PhysRevLett.116.200601}
}

@article{Zhang,
doi = {10.1209/0295-5075/122/17002},
url = {https://dx.doi.org/10.1209/0295-5075/122/17002},
year = {2018},
month = {may},
publisher = {EDP Sciences, IOP Publishing and Società Italiana di Fisica},
volume = {122},
number = {1},
pages = {17002},
author = {Yanchao Zhang and Zhimin Yang and Xin Zhang and Bihong Lin and Guoxing Lin and Jincan Chen},
title = {Coulomb-coupled quantum-dot thermal transistors},
journal = {Europhysics Letters}
}

@article{Mandarino,
  title = {Thermal Transistor Effect in Quantum Systems},
  author = {Mandarino, Antonio and Joulain, Karl and G\'omez, Melisa Dom\'{\i}nguez and Bellomo, Bruno},
  journal = {Phys. Rev. Appl.},
  volume = {16},
  issue = {3},
  pages = {034026},
  numpages = {15},
  year = {2021},
  month = {Sep},
  publisher = {American Physical Society},
  doi = {10.1103/PhysRevApplied.16.034026},
  url = {https://link.aps.org/doi/10.1103/PhysRevApplied.16.034026}
}

@article{Alicki_Fannes,
  title = {Entanglement boost for extractable work from ensembles of quantum batteries},
  author = {Alicki, Robert and Fannes, Mark},
  journal = {Phys. Rev. E},
  volume = {87},
  issue = {4},
  pages = {042123},
  numpages = {4},
  year = {2013},
  month = {Apr},
  publisher = {American Physical Society},
  doi = {10.1103/PhysRevE.87.042123},
  url = {https://link.aps.org/doi/10.1103/PhysRevE.87.042123}
}

@misc{Campaioli,
      title={Quantum Batteries - Review Chapter}, 
      author={Francesco Campaioli and Felix A. Pollock and Sai Vinjanampathy},
      year={2018},
      eprint={1805.05507},
      archivePrefix={arXiv},
      primaryClass={quant-ph}
}

@article{Clivaz,
  title = {Unifying Paradigms of Quantum Refrigeration: A Universal and Attainable Bound on Cooling},
  author = {Clivaz, Fabien and Silva, Ralph and Haack, G\'eraldine and Brask, Jonatan Bohr and Brunner, Nicolas and Huber, Marcus},
  journal = {Phys. Rev. Lett.},
  volume = {123},
  issue = {17},
  pages = {170605},
  numpages = {6},
  year = {2019},
  month = {Oct},
  publisher = {American Physical Society},
  doi = {10.1103/PhysRevLett.123.170605},
  url = {https://link.aps.org/doi/10.1103/PhysRevLett.123.170605}
}

@article{Mitchison,
author = {Mark T. Mitchison},
title = {Quantum thermal absorption machines: refrigerators, engines and clocks},
journal = {Contemporary Physics},
volume = {60},
number = {2},
pages = {164-187},
year = {2019},
publisher = {Taylor & Francis},
doi = {10.1080/00107514.2019.1631555},


URL = { 
    
        https://doi.org/10.1080/00107514.2019.1631555
    
    

},
eprint = { 
    
        https://doi.org/10.1080/00107514.2019.1631555
    
    

}

}

@article{Scovil,
  title = {Three-Level Masers as Heat Engines},
  author = {Scovil, H. E. D. and Schulz-DuBois, E. O.},
  journal = {Phys. Rev. Lett.},
  volume = {2},
  issue = {6},
  pages = {262--263},
  numpages = {0},
  year = {1959},
  month = {Mar},
  publisher = {American Physical Society},
  doi = {10.1103/PhysRevLett.2.262},
  url = {https://link.aps.org/doi/10.1103/PhysRevLett.2.262}
}

@article{TRay2023,
  title = {Optimal quantum resource generation by coupled transmons immersed in Markovian baths},
  author = {Ray, Tanaya and Ghoshal, Ahana and Rakshit, Debraj and Sen, Ujjwal},
  journal = {Phys. Rev. A},
  volume = {108},
  issue = {5},
  pages = {052417},
  numpages = {12},
  year = {2023},
  month = {Nov},
  publisher = {American Physical Society},
  doi = {10.1103/PhysRevA.108.052417},
  url = {https://link.aps.org/doi/10.1103/PhysRevA.108.052417}
}

@Article{Sanderson1,
AUTHOR = {Sanderson, Conrad and Curtin, Ryan},
TITLE = {Practical Sparse Matrices in C++ with Hybrid Storage and Template-Based Expression Optimisation},
JOURNAL = {Mathematical and Computational Applications},
VOLUME = {24},
YEAR = {2019},
NUMBER = {3},
ARTICLE-NUMBER = {70},
URL = {https://www.mdpi.com/2297-8747/24/3/70},
ISSN = {2297-8747},
DOI = {10.3390/mca24030070}
}

@article{Sanderson2016,
  doi = {10.21105/joss.00026},
  url = {https://doi.org/10.21105/joss.00026},
  year = {2016},
  publisher = {The Open Journal},
  volume = {1},
  number = {2},
  pages = {26},
  author = {Conrad Sanderson and Ryan Curtin},
  title = {Armadillo: a template-based C++ library for linear algebra},
  journal = {Journal of Open Source Software}
}

@article{transient,
  title = {Transient effects in quantum refrigerators with finite environments},
  author = {Bhattacharyya, A. and Ghoshal, A. and Sen, U.},
  journal = {Phys. Rev. A},
  volume = {111},
  issue = {1},
  pages = {012209},
  numpages = {18},
  year = {2025},
  month = {Jan},
  publisher = {American Physical Society},
  doi = {10.1103/PhysRevA.111.012209},
  url = {https://link.aps.org/doi/10.1103/PhysRevA.111.012209}
}

@article{Leghtas2015,
  title = {Confining the state of light to a quantum manifold by engineered two-photon loss},
  volume = {347},
  ISSN = {1095-9203},
  url = {http://dx.doi.org/10.1126/science.aaa2085},
  DOI = {10.1126/science.aaa2085},
  number = {6224},
  journal = {Science},
  publisher = {American Association for the Advancement of Science (AAAS)},
  author = {Leghtas,  Z. and Touzard,  S. and Pop,  I. M. and Kou,  A. and Vlastakis,  B. and Petrenko,  A. and Sliwa,  K. M. and Narla,  A. and Shankar,  S. and Hatridge,  M. J. and Reagor,  M. and Frunzio,  L. and Schoelkopf,  R. J. and Mirrahimi,  M. and Devoret,  M. H.},
  year = {2015},
  month = feb,
  pages = {853–857}
}

@article{Firstenberg2013,
  title = {Attractive photons in a quantum nonlinear medium},
  volume = {502},
  ISSN = {1476-4687},
  url = {http://dx.doi.org/10.1038/nature12512},
  DOI = {10.1038/nature12512},
  number = {7469},
  journal = {Nature},
  publisher = {Springer Science and Business Media LLC},
  author = {Firstenberg,  Ofer and Peyronel,  Thibault and Liang,  Qi-Yu and Gorshkov,  Alexey V. and Lukin,  Mikhail D. and Vuletić,  Vladan},
  year = {2013},
  month = sep,
  pages = {71–75}
}

@article{Winkler2006,
  title = {Repulsively bound atom pairs in an optical lattice},
  volume = {441},
  ISSN = {1476-4687},
  url = {http://dx.doi.org/10.1038/nature04918},
  DOI = {10.1038/nature04918},
  number = {7095},
  journal = {Nature},
  publisher = {Springer Science and Business Media LLC},
  author = {Winkler,  K. and Thalhammer,  G. and Lang,  F. and Grimm,  R. and Hecker Denschlag,  J. and Daley,  A. J. and Kantian,  A. and B\"{u}chler,  H. P. and Zoller,  P.},
  year = {2006},
  month = jun,
  pages = {853–856}
}

@article{PhysRevLett.79.1467,
  title = {Strongly Interacting Photons in a Nonlinear Cavity},
  author = {Imamo\ifmmode \bar{g}\else \={g}\fi{}lu, A. and Schmidt, H. and Woods, G. and Deutsch, M.},
  journal = {Phys. Rev. Lett.},
  volume = {79},
  issue = {8},
  pages = {1467--1470},
  numpages = {0},
  year = {1997},
  month = {Aug},
  publisher = {American Physical Society},
  doi = {10.1103/PhysRevLett.79.1467},
  url = {https://link.aps.org/doi/10.1103/PhysRevLett.79.1467}
}

@article{PhysRevLett.124.213601,
  title = {Supercorrelated Radiance in Nonlinear Photonic Waveguides},
  author = {Wang, Zhihai and Jaako, Tuomas and Kirton, Peter and Rabl, Peter},
  journal = {Phys. Rev. Lett.},
  volume = {124},
  issue = {21},
  pages = {213601},
  numpages = {7},
  year = {2020},
  month = {May},
  publisher = {American Physical Society},
  doi = {10.1103/PhysRevLett.124.213601},
  url = {https://link.aps.org/doi/10.1103/PhysRevLett.124.213601}
}

@article{Gu2017,
  title = {Microwave photonics with superconducting quantum circuits},
  volume = {718–719},
  ISSN = {0370-1573},
  url = {http://dx.doi.org/10.1016/j.physrep.2017.10.002},
  DOI = {10.1016/j.physrep.2017.10.002},
  journal = {Physics Reports},
  publisher = {Elsevier BV},
  author = {Gu,  Xiu and Kockum,  Anton Frisk and Miranowicz,  Adam and Liu,  Yu-xi and Nori,  Franco},
  year = {2017},
  month = nov,
  pages = {1–102}
}

@article{8fp4-15mq,
  title = {Mpemba effect in self-contained quantum refrigerators: Accelerated cooling},
  author = {Mondal, Sayan and Sen, Ujjwal},
  journal = {Phys. Rev. A},
  volume = {113},
  issue = {2},
  pages = {022211},
  numpages = {12},
  year = {2026},
  month = {Feb},
  publisher = {American Physical Society},
  doi = {10.1103/8fp4-15mq},
  url = {https://link.aps.org/doi/10.1103/8fp4-15mq}
}

@article{Alicki2015,
  title = {Non-equilibrium quantum heat machines},
  volume = {17},
  ISSN = {1367-2630},
  url = {http://dx.doi.org/10.1088/1367-2630/17/11/115012},
  number = {11},
  journal = {New J. Phys.},
  publisher = {IOP Publishing},
  author = {Alicki,  Robert and Gelbwaser-Klimovsky,  David},
  year = {2015},
  month = Nov,
  pages = {115012}
}

@article{DeChiara2020,
  title = {Quantum machines powered by correlated baths},
  author = {De Chiara, Gabriele and Antezza, Mauro},
  journal = {Phys. Rev. Res.},
  volume = {2},
  issue = {3},
  pages = {033315},
  numpages = {9},
  year = {2020},
  month = {Aug},
  publisher = {American Physical Society},
  doi = {10.1103/PhysRevResearch.2.033315},
  url = {https://link.aps.org/doi/10.1103/PhysRevResearch.2.033315}
}

@article{deAssis2019,
  title = {Efficiency of a Quantum Otto Heat Engine Operating under a Reservoir at Effective Negative Temperatures},
  author = {de Assis, Rog\'erio J. and de Mendon\ifmmode \mbox{\c{c}}\else \c{c}\fi{}a, Taysa M. and Villas-Boas, Celso J. and de Souza, Alexandre M. and Sarthour, Roberto S. and Oliveira, Ivan S. and de Almeida, Norton G.},
  journal = {Phys. Rev. Lett.},
  volume = {122},
  issue = {24},
  pages = {240602},
  numpages = {5},
  year = {2019},
  month = {Jun},
  publisher = {American Physical Society},
  doi = {10.1103/PhysRevLett.122.240602},
  url = {https://link.aps.org/doi/10.1103/PhysRevLett.122.240602}
}

@article{Klaers2017,
  title = {Squeezed Thermal Reservoirs as a Resource for a Nanomechanical Engine beyond the Carnot Limit},
  author = {Klaers, Jan and Faelt, Stefan and Imamoglu, Atac and Togan, Emre},
  journal = {Phys. Rev. X},
  volume = {7},
  issue = {3},
  pages = {031044},
  numpages = {6},
  year = {2017},
  month = {Sep},
  publisher = {American Physical Society},
  doi = {10.1103/PhysRevX.7.031044},
  url = {https://link.aps.org/doi/10.1103/PhysRevX.7.031044}
}

@article{Kirchmair2013,
  title = {Observation of quantum state collapse and revival due to the single-photon Kerr effect},
  volume = {495},
  ISSN = {1476-4687},
  url = {http://dx.doi.org/10.1038/nature11902},
  DOI = {10.1038/nature11902},
  number = {7440},
  journal = {Nature},
  publisher = {Springer Science and Business Media LLC},
  author = {Kirchmair,  Gerhard and Vlastakis,  Brian and Leghtas,  Zaki and Nigg,  Simon E. and Paik,  Hanhee and Ginossar,  Eran and Mirrahimi,  Mazyar and Frunzio,  Luigi and Girvin,  S. M. and Schoelkopf,  R. J.},
  year = {2013},
  month = Mar,
  pages = {205}
}

@article{He2023,
  title = {Fast generation of Schr\"{o}dinger cat states using a Kerr-tunable superconducting resonator},
  volume = {14},
  ISSN = {2041-1723},
  url = {http://dx.doi.org/10.1038/s41467-023-42057-0},
  number = {1},
  journal = {Nat. Commun.},
  publisher = {Springer Science and Business Media LLC},
  author = {He,  X. L. and Lu,  Yong and Bao,  D. Q. and Xue,  Hang and Jiang,  W. B. and Wang,  Z. and Roudsari,  A. F. and Delsing,  Per and Tsai,  J. S. and Lin,  Z. R.},
  year = {2023},
  month = Oct 
}

@article{Miano2022,
  title = {Frequency-tunable Kerr-free three-wave mixing with a gradiometric SNAIL},
  volume = {120},
  ISSN = {1077-3118},
  url = {http://dx.doi.org/10.1063/5.0083350},
  number = {18},
  journal = {Appl. Phys. Lett.},
  publisher = {AIP Publishing},
  author = {Miano,  A. and Liu,  G. and Sivak,  V. V. and Frattini,  N. E. and Joshi,  V. R. and Dai,  W. and Frunzio,  L. and Devoret,  M. H.},
  year = {2022},
  pages = {184002}
}

@article{Bothner2022,
  title = {Four-wave-cooling to the single phonon level in Kerr optomechanics},
  volume = {5},
  ISSN = {2399-3650},
  url = {http://dx.doi.org/10.1038/s42005-022-00808-3},
  number = {1},
  journal = {Commun. Phys.},
  publisher = {Springer Science and Business Media LLC},
  author = {Bothner,  Daniel and Rodrigues,  Ines C. and Steele,  Gary A.},
  year = {2022},
  pages = {33}
}

@article{Zoepfl2023,
  title = {Kerr Enhanced Backaction Cooling in Magnetomechanics},
  author = {Zoepfl, D. and Juan, M. L. and Diaz-Naufal, N. and Schneider, C. M. F. and Deeg, L. F. and Sharafiev, A. and Metelmann, A. and Kirchmair, G.},
  journal = {Phys. Rev. Lett.},
  volume = {130},
  issue = {3},
  pages = {033601},
  numpages = {6},
  year = {2023},
  month = {Jan},
  publisher = {American Physical Society},
  doi = {10.1103/PhysRevLett.130.033601},
  url = {https://link.aps.org/doi/10.1103/PhysRevLett.130.033601}
}

@article{DiazNaufal2025,
  title = {Kerr-enhanced optomechanical cooling in the unresolved-sideband regime},
  author = {Diaz-Naufal, N. and Deeg, L. and Zoepfl, D. and Schneider, C. M. F. and Juan, M. L. and Kirchmair, G. and Metelmann, A.},
  journal = {Phys. Rev. A},
  volume = {111},
  issue = {5},
  pages = {053505},
  numpages = {15},
  year = {2025},
  month = {May},
  publisher = {American Physical Society},
  doi = {10.1103/PhysRevA.111.053505},
  url = {https://link.aps.org/doi/10.1103/PhysRevA.111.053505}
}

@article{Hyyppa2022,
  title = {Unimon qubit},
  volume = {13},
  ISSN = {2041-1723},
  url = {http://dx.doi.org/10.1038/s41467-022-34614-w},
  number = {1},
  journal = {Nat. Commun.},
  publisher = {Springer Science and Business Media LLC},
  author = {Hyypp\"{a},  Eric and Kundu,  Suman and Chan,  Chun Fai and Gunyhó,  András and others 
  },
  year = {2022},
    pages = {6895}
}

@article{Hofheinz2009,
  title = {Synthesizing arbitrary quantum states in a superconducting resonator},
  volume = {459},
  ISSN = {1476-4687},
  url = {http://dx.doi.org/10.1038/nature08005},
  number = {7246},
  journal = {Nature},
  publisher = {Springer Science and Business Media LLC},
  author = {Hofheinz,  Max and Wang,  H. and Ansmann,  M. and Bialczak,  Radoslaw C. and Lucero,  Erik and Neeley,  M. and O’Connell,  A. D. and Sank,  D. and Wenner,  J. and Martinis,  John M. and Cleland,  A. N.},
  year = {2009},
  month = May,
  pages = {546}
}
\end{document}